\documentclass[12pt]{article}

\usepackage[T1]{fontenc}
\usepackage[utf8]{inputenc}
\usepackage{amsmath,amssymb,slashed}
\usepackage{amsfonts}
\usepackage{upgreek}
\usepackage{comment}

\numberwithin{equation}{section}

\usepackage[english]{babel}

\usepackage[mathscr]{eucal}
\usepackage{epsfig}
\usepackage{booktabs}
\usepackage{bbold}
\usepackage{slashed}

\DeclareMathAlphabet{\pxitfont}{OML}{pxmi}{m}{it}
\DeclareMathAlphabet{\pxitfontn}{U}{pxmia}{m}{it}

\DeclareFontFamily{U}{euc}{}%
\DeclareFontShape{U}{euc}{m}{n}{<-6>eurm5<6-8>eurm7<8->eurm10}{}%
\DeclareSymbolFont{AMSc}{U}{euc}{m}{n} %
\DeclareMathSymbol{\psitt}{\mathord}{AMSc}{"20}    
\DeclareMathSymbol{\chitt}{\mathord}{AMSc}{"1F}    

\def\eff{{\mathrm{eff}}}
\def\h{\widehat}

\def\bp{\begin{pmatrix}}
\def\ep{\end{pmatrix}}
\def\sR{{\sf R}}
\def\sT{{\sf T}}
\def\sP{{\sf P}}
\def\sPT{{\sf{PT}}}
\def\HH{{\mathfrak H}}
\def\ICS{{\mathrm{CS}}}
\newcommand{\bea}{\begin{array}}
\newcommand{\eea}{\end{array}}
\newcommand{\beq}{\begin{equation}}
\newcommand{\eeq}{\end{equation}}
\newcommand{\beqn}{\begin{eqnarray}}
\newcommand{\eeqn}{\end{eqnarray}}

\newcommand{\Tr}{{\rm Tr}}

\newcommand{\B}{\mathcal B}

\newcommand{\eps}{\epsilon}
\newcommand{\veps}{\varepsilon}

\renewcommand{\L}{{\mathcal L}}

\newcommand{\RR}{{\mathbf R}}

\newcommand{\Z}{\ZZ}

\def\sign{{\rm sign}\,}

\def\F{{\mathcal F}}
\def\d{{\mathrm d}}
\def\C{{\mathbf C}}
\def\Gr{{\mathrm{Gr}}}

\def\g{\gamma}

\font\teneusm=eusm10 
\font\seveneusm=eusm7 
\font\fiveeusm=eusm5
\newfam\eusmfam
\textfont\eusmfam=\teneusm \scriptfont\eusmfam=\seveneusm
\scriptscriptfont\eusmfam=\fiveeusm

\font\tencmmib=cmmib10 \skewchar\tencmmib='177
\font\sevencmmib=cmmib7 \skewchar\sevencmmib='177
\font\fivecmmib=cmmib5 \skewchar\fivecmmib='177
\newfam\cmmibfam
\textfont\cmmibfam=\tencmmib \scriptfont\cmmibfam=\sevencmmib
\scriptscriptfont\cmmibfam=\fivecmmib

\font\teneurm=eurm10 
\font\seveneurm=eurm7 
\font\fiveeurm=eurm5
\newfam\eurmfam
\textfont\eurmfam=\teneurm \scriptfont\eurmfam=\seveneurm
\scriptscriptfont\eurmfam=\fiveeurm

\font\teneufm=eufm10 
\font\seveneufm=eufm7 
\font\fiveeufm=eufm5
\newfam\eufmfam
\textfont\eufmfam=\teneufm \scriptfont\eufmfam=\seveneufm
\scriptscriptfont\eufmfam=\fiveeufm

\def\H{{\mathcal H}}

\def\bar{\overline}

\def\hat{\widehat}

\def\A{{\mathcal A}}
\def\Z{{\mathbf Z}}
\def\i{{\mathrm i}}

\def\AA{{\mathcal A}}

\def\d{{\mathrm d}}

\def\2{{\bf 2}}
\def\1{{\bf 1}}
\def\0{{\bf 0}}

\def\bar{\overline}

\def\O{{\mathcal O}}

\def\be{\begin{equation}}
\def\ee{\end{equation}}
\def\g{\gamma}

\begin{document}

\thispagestyle{empty}
\begin{flushright}\footnotesize
~

\vspace{2.1cm}
\end{flushright}

\begin{center}
{\Large\textbf{\mathversion{bold} Three Lectures On Topological Phases Of Matter}\par}

\vspace{2.1cm}

\textrm{ Edward Witten}

\vskip1cm

\textit{ School
of Natural Sciences, Institute for Advanced Study, Princeton, NJ 08540}\\
 \vspace{3mm}

\par\vspace{2cm}

\textbf{Abstract}
\end{center}
\noindent  
These notes are based on lectures at the PSSCMP/PiTP summer school that was held at Princeton University and the Institute for Advanced Study
in July, 2015.  They are devoted largely to topological phases of matter that can be understood  in terms of free fermions and band theory.
They also contain an introduction to the fractional quantum Hall effect from the point of view of effective field theory.
\vspace*{\fill}

\setcounter{page}{1}

\newpage

\tableofcontents

\vskip2cm
\setcounter{section}{-1}
\section{Introduction}

In recent years, a number of fascinating new applications of quantum field theory in condensed matter physics have been discovered.  
For an entr\'{e}e to the literature, see the review articles  \cite{KH,ZQ,MZ} and the book \cite{BH}.  

The present notes are based on the first three of four lectures that I gave on these matters  at 
the PSSCMP/PiTP summer school at Princeton University
and the Institute for Advanced Study in July, 2015.  
These lectures contained very little novelty; I simply explained what I have been able to understand of a fascinating subject.
(The fourth lecture did contain some novelty and has been written up separately \cite{Witten}.)  The references include some classic
recent and less recent papers, but they are certainly not complete.

In these lectures,
I mostly concentrated on phases of matter
that can be understood in terms of noninteracting electrons and topological band theory.  
The main exception was a short introduction to
some aspects of the fractional quantum Hall effect.

These notes mostly follow the original lectures rather closely.  Some topics have been slightly rearranged and
a few matters for which unfortunately there was no time in the original lectures have been added.  Some topics treated here were described
from a different point of view by other lecturers at the school, especially Charlie Kane and Nick Read.

\section{Lecture One}

\subsection{Relativistic Dispersion In One Space Dimension}

\begin{figure}
 \begin{center}
   \includegraphics[width=3in]{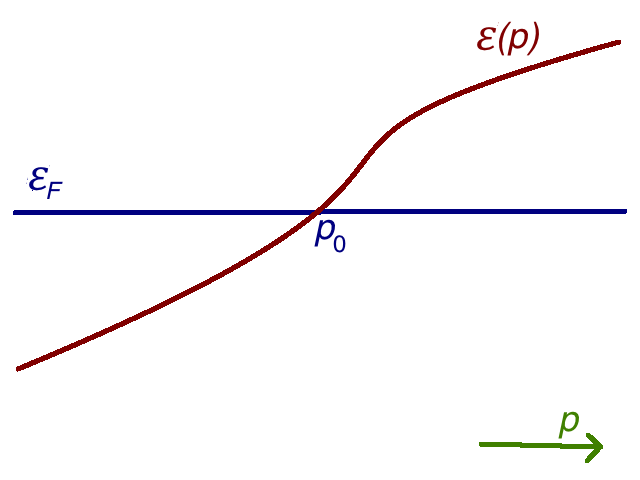}
 \end{center}
\caption{\small  In one dimension, the single-particle energy $\veps(p)$ generically crosses the Fermi energy at an isolated
momentum $p_0$.  }
 \label{cigar}
\end{figure}

We will start by asking under what conditions we should expect to find a relativistic dispersion relation for electrons in a crystal.
In one space dimension, the answer is familiar.  Writing $\veps(p)$ for the single particle energy 
$\veps$ as a function of momentum $p$,
generically $\veps(p)$ crosses the Fermi energy with a nonzero slope at some $p=p_0$ (fig. \ref{cigar}).

Then
linearizing the dispersion relation around $p=p_0$, we get
\begin{equation}\label{dolz}\veps=\veps(p_0)+v(p-p_0)+\O((p-p_0)^2),~~~ v=\left.\frac{\partial \veps}{\partial p}\right|_{p=p_0}.\end{equation}
Apart from the additive constant $\veps(p_0)$ and the shift $p\to p-p_0$,
this  is a relativistic dispersion relation, analogous to $\veps=cp$,  with the speed of light $c$ replaced by $v$.    For $v>0$ ($v<0$), the
gapless mode that lives near $p=p_0$ travels to the right (left).

The corresponding continuum model describing the modes near $p=p_0$ is
\begin{equation}\label{continuum} H=v\int_{-\infty}^\infty \d x \,\,\psi^*\left(-\i\frac{\partial}{\partial x}\right)\psi.\end{equation}  
This is a relativistic action for a 1d chiral fermion, except that $v$ appears instead of $c$ and $-i\partial/\partial x$
represents $p-p_0$ instead of $p$.  Also we have omitted from $H$
the ``constant'' $\veps(p_0)$ per particle:
\be\label{constantshift}\veps(p_0)\int_{-\infty}^\infty \d x \,\psi^*\psi.\ee

\begin{figure}
 \begin{center}
   \includegraphics[width=3in]{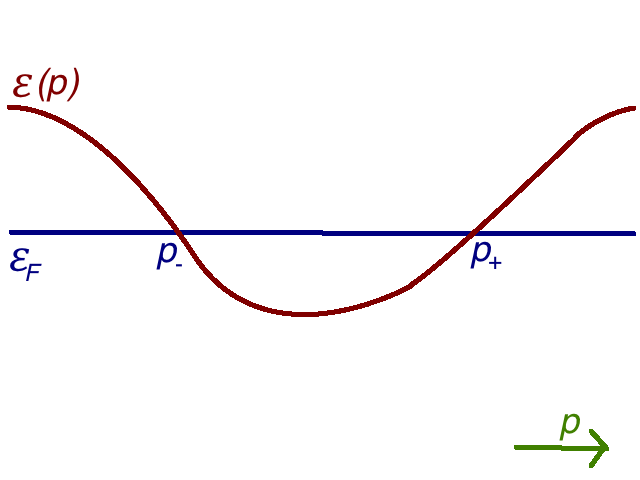}
 \end{center}
\caption{\small  In one dimension, for every value of the momentum at which  $\veps(p)$ increases above $\eps_F$,
there is another point at which it decreases below $\veps_F$. }
 \label{Crossings}
\end{figure}

This one-dimensional case gives an easy first example of how global conditions in topology constrain the possible low
energy field theory that we can get -- and how these constraints often mirror familiar facts about relativistic field theory and ``anomalies.''  
We have to remember that in the context of a crystal, the momentum $p$ is a periodic variable.
Because $\veps(p)$ is periodic,  it follows (fig. \ref{Crossings}) that for every time $\veps(p)$ crosses the
Fermi energy $\veps_F$ from below, there is another time that it crosses $\veps_F$ from above
 So actually there are equally many gapless left-moving and right-moving fermion modes.
 
 In relativistic terminology, the right-moving
and left-moving modes are said to have positive and negative chirality.   The motivation for this terminology is that the massless Dirac equation in 2 spacetime dimensions 
is \be\label{Dirac}\left(\gamma^0\frac{\partial}{\partial t}+\gamma^1\frac{\partial}{\partial x}\right)\psi=0,\ee
where $\g^\mu$ are Dirac matrices, obeying the Clifford algebra relations
\be\label{gamma}\{\gamma^\mu,\gamma^\nu\}=2\eta^{\mu\nu},~~~~\eta_{\mu\nu}=\mathrm{diag}(-1,1).\ee
   In Hamiltonian form, the Dirac equation is
\be\label{hamdir}\i\frac{\partial\psi}{\partial t}= -\i\bar\gamma \frac{\partial\psi}{\partial x},\ee
where
$\bar\gamma=\gamma_0\gamma_1$
(whose analog in $3+1$ dimensions is usually called $\gamma_5$) is the ``chirality operator.''    So a fermion state of positive or negative
chirality is right-moving or left-moving.

Thus a more realistic Hamiltonian for the gapless charged modes will be something like
\be\label{realism}H=-v_-\int_{-\infty}^\infty \d x \,\,\psi_-^*\left(-\i\frac{\partial}{\partial x}\right)\psi_-+ 
v_+\int_{-\infty}^\infty \d x \,\,\psi_+^*\left(-\i\frac{\partial}{\partial x}\right)\psi_+,\ee
where $\psi_+$ and $\psi_-$ are modes of positive and negative chirality, and in general they propagate with different
velocities.
 If one is familiar with quantum gauge theories and anomalies, one will recognize that this topological fact -- which is a 1d analog of the 3d Nielson-Ninomiya
theorem that we get to presently -- has saved us from trouble.   A purely $1+1$-dimensional theory with, say, 
$n_+$ right-moving gapless electron modes and $n_-$
left-moving ones is ``anomalous,'' meaning that it is not gauge-invariant and does not conserve electric charge -- unless $n_+=n_-$.  The
anomaly is the $1+1$-dimensional version of the Adler-Bell-Jackiw anomaly  \cite{Adler,BJ}, which is very important in particle physics.

We can actually see the potential anomaly by re-examining fig. \ref{Crossings},  but now assuming that a constant electric field is turned on.  In the presence of an electric field with a sign
such that $\d p/\d t>0$ for each electron, the electrons will all ``flow'' to the right in the picture.   This creates electrons at $p=p_+$ and holes
at $p=p_-$,  so the charge carried by the $p=p_+$ mode
or by the $p=p_-$ mode is not conserved, although the total charge is conserved, of course.   Thus charge conservation depends on having both
types of mode equally.

\subsection{Three Dimensions}

\begin{figure}
 \begin{center}
   \includegraphics[width=2.5in]{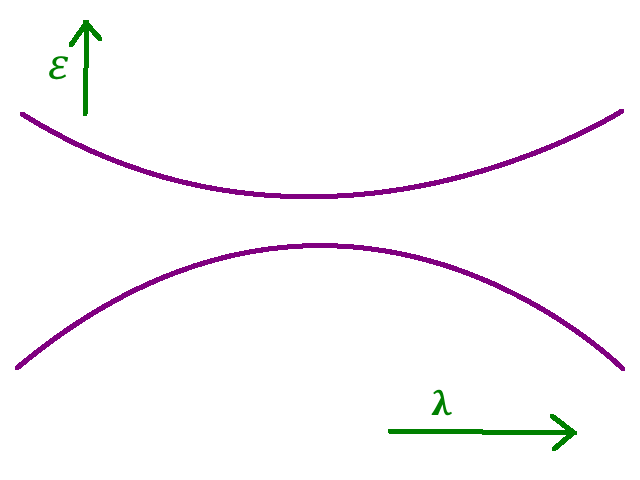}
 \end{center}
\caption{\small  Generically, quantum mechanical energy levels do not cross as a parameter is varied. }
 \label{repel}
\end{figure}

There is certainly more that one could say in 1 space dimension, but instead we are going to go on to spatial dimension 3.    As a preliminary, recall that quantum mechanical
energy levels repel, which means that if $H(\lambda)$ is a generic 1-parameter family of Hamiltonians, depending on a parameter $\lambda$, and with no particular
symmetry, then generically its energy levels do not cross as a function of $\lambda$ (fig. \ref{repel}).
But \cite{Herring} how much do levels repel each other?  Generically, how many parameters do we have to adjust to make two energy levels coincide?

The answer to this question is that we have to adjust 3 real parameters,  because a generic $2\times 2$ Hermitian matrix depends on 4 real parameters
\be\label{generic} H=\begin{pmatrix} a & b \cr \bar b & c\end{pmatrix} , \ee
but a $2\times 2$ Hermitian matrix whose energy levels are equal depends on only 1 real parameter
\be\label{specil}H = \begin{pmatrix} a & 0 \cr 0 & a \end{pmatrix}.\ee
  To put this differently, any $2\times 2$ Hermitian matrix is
\be\label{ecial}H = a+\vec b\cdot \vec \sigma,\ee
where $\vec\sigma$ are the Pauli matrices.   The condition for $H$ to have equal eigenvalues is $\vec b=0$, and this is three real conditions.

In three dimensions, a band Hamiltonian $H(p_1,p_2,p_3)$ depends on three real parameters, so it is natural for two bands to cross at some
isolated value $p=p_*$.      Near $p=p_*$, and looking only at the two bands in question, the Hamiltonian looks something like
\be\label{localexp}H=a( p)+ \vec b( p)\cdot \vec\sigma,\ee
where $\vec b(p)=0$ at $p=p_*$.  Expanding near $p=p_*$,
\be\label{expa}b_i(p)=\sum_j b_{ij} (p-p_*)_j+\O((p-p_*)^2),~~ b_{ij}= \left.\frac{\partial b_i}{\partial p_j}\right|_{p=p_*}.\ee  Thus dropping a constant
and ignoring
higher order terms, the band splitting is described near $p=p_*$ by
\be\label{xpa}H'=\sum_{i,j}\sigma_i b_{ij}(p-p_*)_j.\ee

Apart from a shift $p\to p-p_*$, this is essentially a chiral Dirac Hamiltonian in $3+1$ dimensions.  Let us review this fact.   The massless
Dirac equation in $3+1$ dimensions is 
\be\label{thred}\sum_{\mu=0}^3\gamma^\mu\partial_\mu\psi=0,~~~ \{\gamma^\mu,\gamma^\nu\}=2\eta^{\mu\nu},~~~~\eta_{\mu\nu}=\mathrm{diag}(-1,1,1,1).\ee
In Hamiltonian form, this equation is
\be\label{hameq}\i\frac{\partial\psi}{\partial t}=-\i\sum_k\gamma_0\gamma_k \frac{\partial\psi}{\partial x^k}.  \ee
  To represent the four gamma matrices, we need $4\times 4$ matrices (which can be chosen to be real). 
However, the matrix
\be\label{chirm}\gamma_5=\i\gamma_0\gamma_1\gamma_2\gamma_3\ee is Lorentz-invariant.   It obeys $\gamma_5^2=1$, so its eigenvalues are $\pm 1$.
We can place on $\psi$ a ``chirality condition'' $\gamma_5\psi=\pm \psi$,  reducing to a $2\times 2$ Dirac Hamiltonian. But then, because of the factor of $\i$ in the definition of $\gamma_5$, and in
contrast to what happens in $1+1$ dimensions, the adjoint of $\psi$ obeys the opposite chirality condition.  

Once we reduce to a chiral $2\times 2$ Dirac Hamiltonian with $\gamma_5\psi=\pm\psi$, the matrices
$\gamma_0\gamma_i$ that appear in the Dirac Hamiltonian are $2\times 2$ hermitian matrices and we can take them to be, up to sign, the Pauli
sigma matrices
\be\label{sigmam}\sigma_i=\pm \gamma_0\gamma_i.\ee
 The point  is that, if $\gamma_5\psi=\pm \psi$,
then {\it in acting on $\psi$},
\be\label{news}\sigma_i\sigma_j=\delta_{ij}+\i\epsilon_{ijk}\sigma_k.\ee
(One may prove this for $i=1,j=2$ from the explicit identity
$\gamma_0\gamma_1\gamma_0\gamma_2=\i\gamma_0\gamma_3\gamma_5$. The general case then follows from
rotation symmetry.)

\begin{figure}
 \begin{center}
   \includegraphics[width=2.5in]{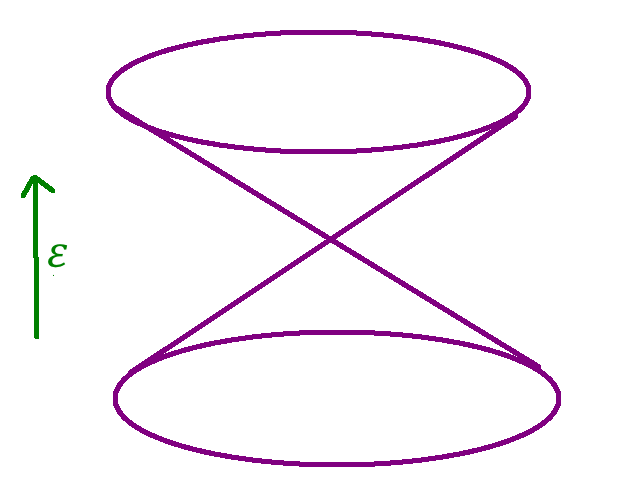}
 \end{center}
\caption{\small  A pair of bands described by a chiral Dirac Hamiltonian. }
 \label{twob}
\end{figure}

So the Dirac Hamiltonian
\be\label{dirha}  H=-\i\sum_k\gamma_{0}\gamma_k\frac{\partial}{\partial x^k} \ee
becomes for a chiral fermion
\be\label{irha} H=\mp \i c\sum_k\sigma^k\frac{\partial}{\partial x^k}=\pm c\,\vec\sigma\cdot \vec p.\ee   
The sign depends on the fermion chirality, which determines which sign we had to pick in eqn. (\ref{sigmam}).
 (I have restored $c$, the speed of light.)    As a matter of terminology, a charged relativistic fermion of
definite chirality -- in other words, with a definite value of $\gamma_5$ -- is called a Weyl fermion.   The physical meaning of the eigenvalue of $\gamma_5$
is that it determines the fermion ``helicity'' (spin around the direction of motion).  Note that ``helicity'' is only a Lorentz-invariant
notion  for a {\it massless} particle
(which is never at rest in any Lorentz frame) 
and indeed our starting point was
 the massless Dirac equation.   The {\it antiparticle} -- which one can think of as a hole in the Dirac sea --
has opposite helicity,\footnote{This happens because -- in contrast to what happens in $1+1$ dimensions --
if the $\gamma^\mu$ are real then the chirality operator $\g_5$ is imaginary.  Accordingly $\psi$ and its hermitian adjoint obey opposite chirality
conditions.  The basic example in particle physics is that, in the approximation in which they are massless, neutrinos and antineutrinos have
opposite helicity.}  
 somewhat as it has opposite charge.  The chiral Dirac Hamiltonian of eqn. (\ref{irha}) describes two bands with $E=\pm c|p|$ (fig. \ref{twob}).

Thus, the chiral Dirac Hamiltonian basically coincides with the generic Hamiltonian (\ref{xpa}) that we found for a $2\times 2$
band crossing,
with the replacement $cp_k\to \sum_j b_{kj}p_j$.    This replacement means, of course, that the fermion modes near $p=p_*$ do
not propagate at velocity $c$ but much more slowly.  Also, they do not necessarily propagate isotropically in the standard Euclidean metric on $\RR^3$.  In general,
the natural metric governing these modes is
$$||p||^2=\sum_i\left(\sum_j b_{ij}p_j\right)^2.$$ 
In other words, the effective metric is 
$$G^{ij}=\sum_k b^i{}_kb^j{}_k.$$

Finally, and very importantly, the {\it chirality} of the gapless electron mode is given by
\be\label{perd}\mathrm{sign}\,\det\left(b_{ij}\right)=\mathrm{sign}\,\det\left. \left(\frac{\partial b_i}{\partial p_j}\right)\right|_{p=p_*}. \ee
  A gap crossing in which this determinant is positive (or negative) corresponds to a relativistic massless chiral fermion (or Weyl fermion) with
$\gamma_5=+1$ (or $\gamma_5=-1$).

 \subsection{The Nielsen-Ninomiya Theorem}
 
 Now, however, we should remember something about relativistic quantum field theory in $3+1$ dimensions.   A theory of a $U(1)$ gauge
field (of electromagnetism) coupled to a massless chiral charged fermion of one chirality, with no counterpart of the opposite 
chirality, is anomalous: gauge invariance fails at the quantum level, and the theory is inconsistent.
  In $1+1$ dimensions, we avoided such a contradiction because of a simple topological fact that $\veps(p)$ passes downward
through the Fermi  energy as often as it passes upwards, as in fig. \ref{Crossings}.  An analogous topological theorem
 saves the day in $3+1$ dimensions.  This is the Nielsen-Ninomiya theorem \cite{NN,F}, which was originally formulated as an obstruction
to a lattice regularization of relativistic chiral fermions.\footnote{The Nielsen-Ninomiya theorem involves
 ideas somewhat analogous to those developed by Thouless, Kohmoto, Nightingale, and den Nijs \cite{TKNN} in celebrated work on the quantum Hall effect.
 We will describe their result in Lecture Two.}

\begin{figure}
 \begin{center}
   \includegraphics[width=2.5in]{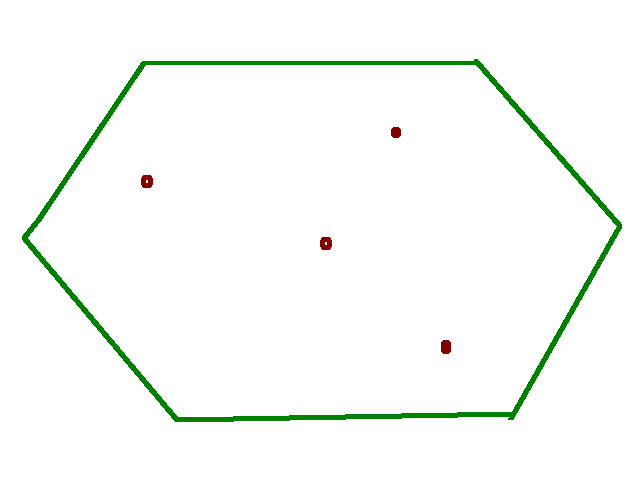}
 \end{center}
\caption{\small  The interior of the hexagon symbolizes the Brillouin zone $\B$.  We consider a band Hamiltonian that is gapped
except at finitely many points, which are indicated by dots. }
 \label{badp}
\end{figure}

In formulating this theorem, we assume that the band Hamiltonian $H(p)$ is gapped except at finitely many isolated points in the Brillouin zone  $\B$ (fig. \ref{badp}).
  We will attach an integer to each of these bad points, and show that these integers add up to 0.

To get started, we assume there are only two bands.   Also, by simply subtracting a $c$-number function of $p$ from  $H(p)$, we can
make $H(p)$  traceless, without changing the band crossings.    
So 
$$H(p)=\vec b(p) \cdot \vec \sigma$$
for some vector-valued function $\vec b(p)$.    Now away from the bad points, $\vec b(p)\not=0$ and so we can define a unit
vector 
$$\vec n(p) =\frac{\vec b}{|\vec b|}. $$
  The mapping $p\to \vec n(p)$ is defined away from the bad points.  We want to understand its topological properties.
  
\begin{figure}
 \begin{center}
   \includegraphics[width=3in]{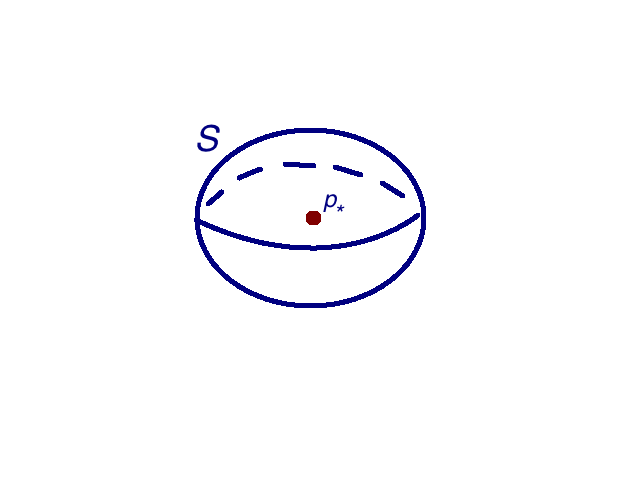}
 \end{center}
\caption{\small  A small sphere $S$ around a bad point $p_*$ at which two levels cross. }
 \label{sphere}
\end{figure}

  Let us consider just one of the bad points, say at $p=p_*$, and let $S$ be a small sphere around this bad point (fig. \ref{sphere}). 
  The map $p\to \vec n(p)$ is defined everywhere on $S$.     This is a mapping from one two-sphere -- namely $S$ -- to another two-sphere --
parametrized by the unit vector $\vec n$. We will call that second two-sphere $S_{\vec n}$.   In any dimension $d$, a continuous mapping from one $d$-dimensional sphere $S^d $ to another sphere of the same dimension always has
a ``winding number'' or ``wrapping number,'' the net number of times the first sphere wraps around the second. 
This reflects the fact that
$$\pi_d(S^d)\cong\Z.$$

Before developing any general theory, let us see what the winding number is in the case of the relativistic Dirac Hamiltonian
$H=\pm \vec \sigma\cdot \vec p,$
where the sign is the fermion chirality.
 For this Hamiltonian, $\vec b=\pm \vec p$, and hence $\vec n=\pm \vec p/|\vec p|$.   The bad point is $\vec p=0$, and we can take the sphere $S$
that surrounds the bad point to be the unit sphere $|\vec p|=1$.  Thus the map from $S$ to $S_{\vec n}$  is just
\be\label{zob}\vec n=\pm \vec p.\ee 
This is the identity map, of winding number 1, in the case of $+$ chirality, and it is minus the identity map, which winds around in 
reverse, with winding
number $-1$, in the case of $-$ chirality.

The Nielsen-Ninomiya theorem is the statement that the sum of the winding numbers at the bad points is always 0.  Generically
(in the absence of lattice symmetries that would lead to a more special behavior) a bad point of winding number bigger than 1 in absolute value
will split into several bad points of winding number $\pm 1$. (We give in section \ref{somex} an explicit example of how this occurs.)  So generically, the bad points all have winding numbers $\pm 1$, corresponding
to gapless Weyl fermions of one chirality or the other. 
   In this case, the vanishing of the sum of the winding numbers means that there
are equally many gapless modes of positive or negative chirality, as a relativistic field theorist would expect for anomaly cancellation.

How does one prove that the sum of the winding numbers is 0?    One rather down-to-earth method is as follows. 
The winding number for a map from $S$ to $S_{\vec n}$ can be expressed as an integral formula:
\be\label{wn}w(S)
=\frac{1}{4\pi}\int\d^2p\,\,\epsilon^{\mu\nu}\vec n\cdot \partial_\mu \vec n\times \partial_\nu \vec n.\ee
An equivalent way to write the same formula is
\be\label{nwn}w(S)=\frac{1}{4\pi}\int_S\d^2 p\,\,\epsilon^{\mu\nu} \,\,\epsilon^{abc}n_a \frac{\partial n_b}{\partial p^\mu}\frac{\partial n_c}{\partial p^\nu}.\ee
  Now 
\be\label{zolf} 0=\partial_\lambda\left(\epsilon^{\lambda\mu\nu} \vec n\cdot \partial_\mu \vec n\times \partial_\nu \vec n\right),\ee since the right hand side
is $\epsilon^{\lambda\mu\nu}\partial_\lambda \vec n\cdot \partial_\mu \vec n\times \partial _\nu \vec n$, which vanishes because it is the triple
cross product of three vectors
 $\partial_\lambda\vec n$, $\partial_\mu \vec n$, and $\partial_\nu\vec n$ that are all normal to the sphere $|\vec n|=1$.

\begin{figure}
 \begin{center}
   \includegraphics[width=2.5in]{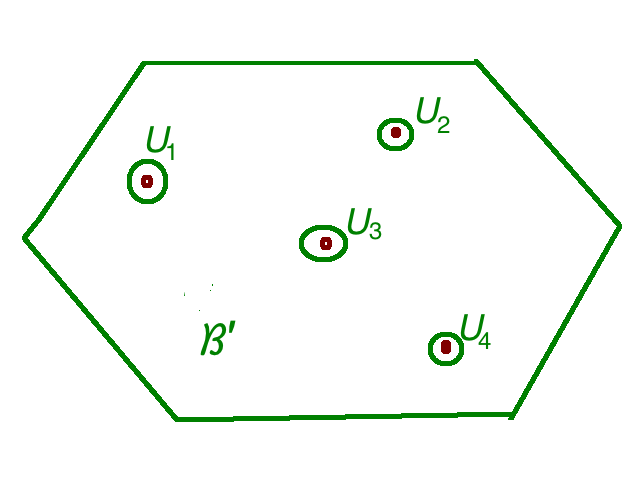}
 \end{center}
\caption{\small $\B'$ is defined by removing a small open set $U_\alpha$ around each bad point $p_\alpha$ in the Brillouin zone.  }
 \label{badt}
\end{figure}

For each bad point $p_\alpha$, let $U_\alpha$ be a small 
open ball around $p_\alpha$ whose boundary is a sphere $S_\alpha$.
Let $\B$ be the full Brillouin zone, and let $\B'$ be what we get by removing from $\B$ all of the $U_\alpha$.   Thus the boundary of
$\B'$ is $\partial \B'=\cup_\alpha S_\alpha$ (fig. \ref{badt}).
Then from Stokes's theorem,
\begin{align}\notag 0&=\frac{1}{4\pi}\int_{B'}\d^3p\,\, \partial_\lambda\left(\epsilon^{\lambda\mu\nu} (n\cdot  
\partial_\mu n\times\partial_\nu n)\right)\cr &=\sum_\alpha\frac{1}{4\pi}\int_{S_\alpha}\d^2 p\,\,\epsilon^{\mu\nu} \,\,
\vec n\cdot \partial_\mu \vec n\times \partial_\nu \vec n \cr &=\sum_\alpha w(S_\alpha).\end{align}
Thus the sum of the winding numbers at bad points is 0, as promised.

In terms of differential forms, one can express this argument more briefly as follows.    Let $\eta$ be the volume form of $S_{\vec n}$.
Thus, $\eta$ is a closed 2-form whose integral is 1:
\be\label{weta}0=\d \eta,~~~\int_{S_{\vec n}}\eta = 1.\ee
Given a map $\varphi:S\to S_{\vec n}$, the corresponding winding number is
\be\label{wnn} w(S)=\int_S\varphi^*(\eta).\ee  So
\be\label{cl}0=\int_{B'}\varphi^*(\d \eta)=\int_{B'}\d \varphi^*(\eta)=\sum_\alpha\int_{S_\alpha}\varphi^*(\eta)=\sum_\alpha w(S_\alpha).\ee

\subsection{The Berry Connection}

Another approach to the same result involves the {\it Berry connection}, and more fundamentally the line bundle on which the
Berry connection is a connection.  This approach is useful in generalizations.
   For each value of $p$ away from the bad points, the Hamiltonian $H(p)$ has one negative eigenvalue,
so the space of filled fermion states of momentum $p$ is a 1-dimensional complex vector space that I will call $\L_p$.  
The fancy way to describe this situation is to say that as $p$ varies, $\L_p$ varies as the fiber of a complex line bundle $\L$ over the Brillouin zone $\B$.
  A vector in $\L_p$
is a wave function $\psi_p$ that obeys $H(p)\psi_p=-\psi_p$.     We can ask for $\psi_p$ to be normalized, $\langle\psi_p,\psi_p\rangle=1$,  but there is
no natural way to fix the {\it phase} of $\psi_p$.

\begin{figure}
 \begin{center}
   \includegraphics[width=2.5in]{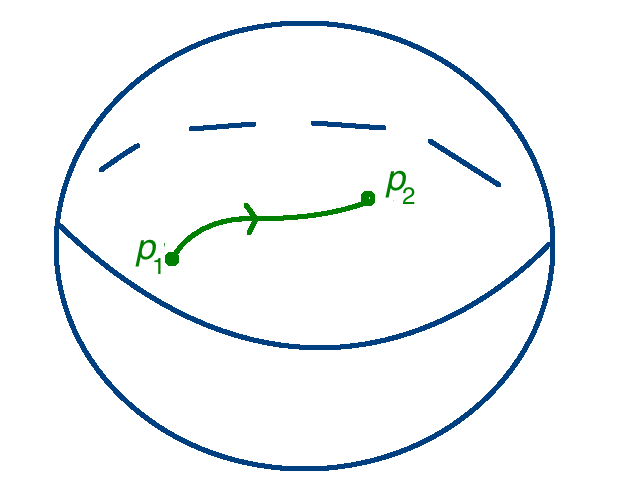}
 \end{center}
\caption{\small Transporting a quantum wavefunction over a path in some parameter space.  One can think of the parameter space as
a sphere that surrounds a bad point in the Brillouin zone.}
 \label{berry}
\end{figure}

 However, suppose that we vary $p$ continuously by a path $p=p(s)$ from, say, $p_1$ to $p_2$, as indicated
in fig. \ref{berry}.  For example, we can consider a  
 path that lies in a sphere $|p-p_*|=\epsilon$ around a bad point $p_*$.  If we make any arbitrary choice of the phase of
$\psi_p$ at $p=p_1$, then we can parallel transport the phase of $\psi_p$ along the given path by requiring that at all $p$ along the path
\be\label{zb}\langle \psi_p,\frac{\d}{\d s}\psi_p\rangle = 0. \ee
Concretely, the real part of this equation ensures that $\langle \psi_p,\psi_p\rangle$ is constant along the path, and the imaginary part of the equation determines
how the phase of $\psi_p$ depends on the parameter $s$.

\begin{figure}
 \begin{center}
   \includegraphics[width=2.5in]{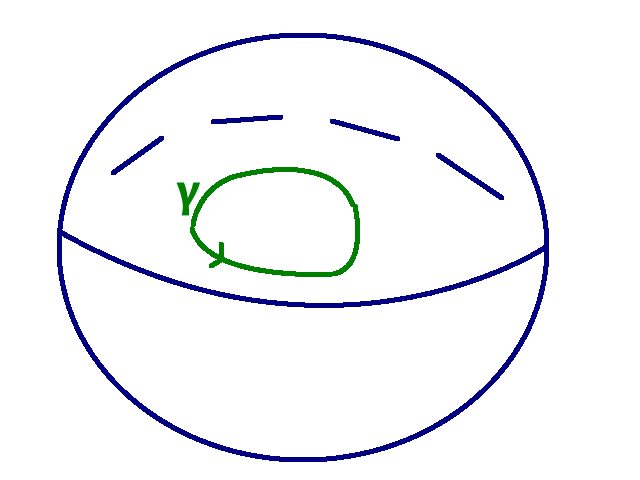}
 \end{center}
\caption{\small Parallel transport of a wavefunction around a closed loop.  }
 \label{closedberry}
\end{figure}

Having a rule of parallel transport of the phase of $\psi_p$ along any path amounts to defining a {\it connection} on $\B'$
(more exactly on the line bundle $\L\to \B'$ whose sections we are parallel transporting).   A {\it connection} on a complex line bundle $\L$
is the same as an abelian gauge field, which we will call $\A$.    Parallel transport around a closed loop $\gamma$ 
using the Berry connection does
not bring us back to the starting point (fig. \ref{closedberry}).  That is, the Berry connection is not flat; it has a curvature  $\F=\d \A$. 
This curvature, divided by $2\pi$, represents (modulo torsion) the {\it first Chern class} of the line bundle $\L\to \B'$:
\be\label{fc}c_1(\L)\longleftrightarrow \frac{\F}{2\pi}.\ee

If $p_\alpha$ is one of the bad points at which two bands cross and $S_\alpha$ is a small sphere around $p_\alpha$, then the flux of
$\F/2\pi$ over the sphere $S_\alpha$ is the winding number, as defined earlier:
\be\label{zif}w_\alpha(S)=\int_{S_\alpha}\frac{\F}{2\pi}.\ee
  The Bianchi identity for any abelian gauge field $\A$ asserts that 
\be\label{wif}\d\F=0.\ee
  So once again we get the Nielsen-Ninomiya theorem
\be\label{nna}0=\int_{B'}\frac{\d \F}{2\pi}=\sum_\alpha\int_{S_\alpha}\frac{\F}{2\pi}=\sum_\alpha w(S_\alpha).\ee

\begin{figure}
 \begin{center}
   \includegraphics[width=2.5in]{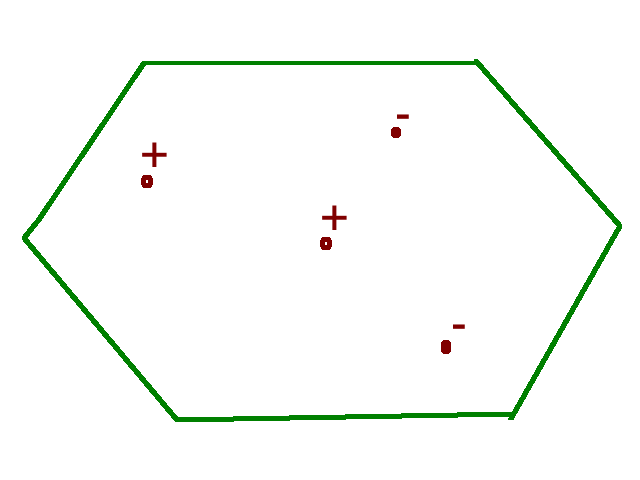}
 \end{center}
\caption{\small Generically, the band-crossing points in the Brillouin zone support  Weyl fermions of positive and negative chirality.  The Nielsen-Ninomiya
theorem says that there are equally many of both types, as shown here. }
 \label{twoob}
\end{figure}

So a more precise picture of the bad points in the Brillouin zone for a generic two-band system is as shown in fig. \ref{twoob}.
A bad point labeled by $+$ or $-$ supports a gapless Weyl fermion of positive or negative chirality; the Nielsen-Ninomiya theorem says that
there are equally many $+$ and $-$ points.

\subsection{Some Examples}\label{somex}

It is instructive to see concretely how two bad points of opposite chirality can annihilate as a parameter is varied.  In relativistic physics,
this can happen as follows (I am jumping ahead slightly, as we have not yet formulated the Nielsen-Ninomiya theorem for a system with more
than two bands).  Consider a four-band system in which the first two bands describe a Weyl fermion of positive chirality and the last
two bands describe a Weyl fermion of negative chirality.  Altogether, these four bands describe a four-component Dirac fermion.  A Dirac
fermion can have a bare mass, and when we add a bare mass term to the Hamiltonian, the band crossings disappear.

This is the usual relativistic picture, but in condensed matter physics, there is more freedom and the annihilation of two bad points can perfectly well occur
for a two-band system.  Consider the explicit Hamiltonian
\be\label{terf}H(p_1,p_2,p_3)=\begin{pmatrix} f(p_3) & p_1-\i p_2\cr p_1+\i p_2 & -f(p_3)\end{pmatrix}.\ee
If $f(p_3)=p_3$, this is the basic Hamiltonian $H=\vec\sigma\cdot \vec p$ of a Weyl fermion.  More generally, if $f(p_3)$ is any smooth
function with only simple zeroes, then a band crossing occurs at any zero of $f(p_3)$ (with $p_1=p_2=0$), and gives a Weyl fermion of positive
or negative chirality depending on the sign of $\d f/\d p_3$ at the zero.  A simple model with
\be\label{werf}f(p_3)=p_3^2-a \ee
gives, for $a>0$, a pair of Weyl points with positive or negative chirality at $p_3=\pm\sqrt a$.  The two Weyl points coalesce for $a=0$ and disappear
for $a<0$.  This phenomenon does not arise in relativistic physics for a two-band system, since the effective Hamiltonian near $\vec p=a=0$ does not
have Lorentz symmetry or even rotation symmetry.  

If is also of some interest to see how a band crossing point of multiplicity $s>1$ can split into $s$ points each of multiplicity 1.
For this, consider the model Hamiltonian
\be\label{teref}H(p_1,p_2,p_3)=\bp p_3 & \bar {g(x)}\,\cr g(x) & -p_3\ep,\ee
where $g$ is a polynomial in the complex variable $x= p_1+\i p_2.$  If $g(x)=x$, we have again the basic Weyl Hamiltonian.
Band crossings occur at zeroes of $g$ (with $p_3=0$).  A simple zero gives a Weyl point of positive chirality and a multiple zero
gives a band crossing point of higher multiplicity.   So for example, the choice $g(x)=\prod_{i=1}^s(x-b_i)$ gives a model with a single
band crossing point of multiplicity $s$ if the $b_i$ are all equal, and $s$ such points each of multiplicity 1 if the $b_i$ are generic.

\subsection{Band Crossing At The Fermi Energy}\label{bce}

\begin{figure}
 \begin{center}
   \includegraphics[width=2.5in]{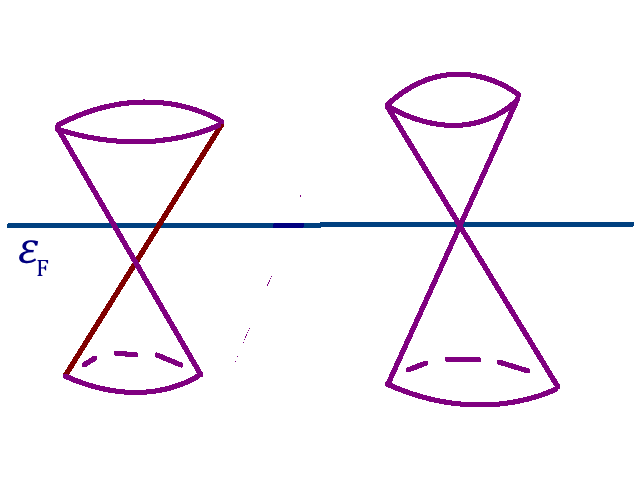}
 \end{center}
\caption{\small Band crossing below the fermi surface (left) or at the fermi surface (right).  }
 \label{choices}
\end{figure}

So far, we have seen how band crossings modeled by a chiral Dirac Hamiltonian can arise naturally in condensed matter physics.
But if we want this to  have striking consequences, it will not do to have the band crossing at a random
energy; we are really only interested in a band crossing that is at, or very near, the Fermi energy $\veps_F$.  Thus we want the picture to look like the one
on the right and not the one on the left in fig. \ref{choices}.  Moreover, it will not do if the band structure is as shown on the right of the figure
in part of the Brillouin zone, and like what is shown on the left in some other part.  In that case, we will get a ``normal metal'' (because of the
band crossing that is above or below $\veps_F$), and its effects will probably swamp the more subtle ``semi-metal'' effects due to the band
crossing which is at $\veps_F$.

\begin{figure}
 \begin{center}
   \includegraphics[width=2.5in]{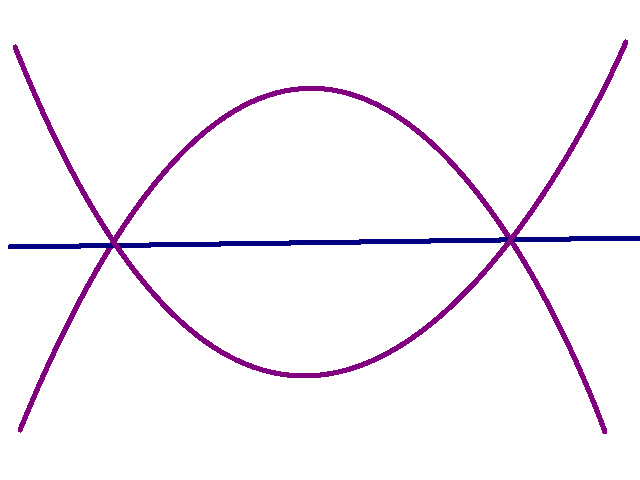}
 \end{center}
\caption{\small Two band-crossing points related by a discrete left-right symmetry.   The fact that the number of electrons per unit cell is an integer
makes it natural for both to occur at the fermi surface.  }
 \label{symbands}
\end{figure}

Ideally, we want the Fermi surface to consist only of a finite set of Weyl points at which two bands cross precisely at $\veps_F$.  There will
have to be an even number of such points, with their chiralities adding to zero.   How can we arrange that {\it all} band crossings occur at $\veps_F$?  As 
a first step, how can we arrange so that they are all at the same
energy?    In the context of condensed matter physics, the way to do this is to find a material that has discrete spatial symmetries (and/or time-reversal symmetries) that permute all of
the bad points.  Some of these symmetries have to be space or time orientation-reversing, since they have to exchange $+$ and $-$ points.    The picture
will then look more like what is shown in fig \ref{symbands}, with a left-right symmetry that exchanges the two bad points.

But how can we arrange so that the energy at which the band crossings occur is precisely $\veps_F$?  Here  we run into one of the beautiful
 things in this subject.   We can get that for free, because the number of electrons per unit cell is an integer.   For example, if there is precisely
 one electron per unit cell that is supposed to be filling the two bands in our model, the Fermi energy will be where we want it, because in fig. \ref{symbands},
 at every value of the momentum $\vec p$ away from the two bad points, precisely one state lies below $\veps_F$ and one lies above $\veps_F$.
 Hence the band-crossing energy is the Fermi energy at half-filling.
 There are many important examples of this phenomenon.  The oldest and best-known is graphene (in two dimensions), which we
 will discuss in Lecture Three.  More recent examples
 involve Weyl semi-metals in three dimensions.  
 
\begin{figure}
 \begin{center}
   \includegraphics[width=2.5in]{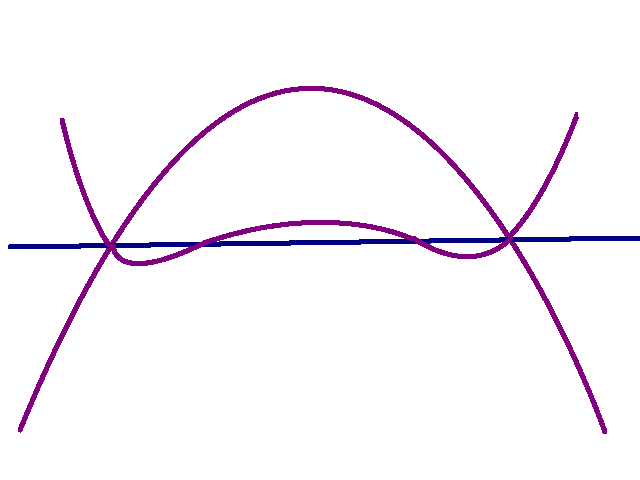}
 \end{center}
\caption{\small This band structure has the same left-right symmetry as in fig. \ref{symbands}, but describes an ordinary metal rather than
a Weyl semimetal.  That is because with this band structure, 
the energy of the band crossings -- indicated by the horizontal line -- passes below some of the states
in the lower band and hence is below the fermi energy.  }
 \label{symbands3}
\end{figure}

 To be more exact, it is natural in this situation to have the band crossings at $\veps_F$ in the sense that, given a band Hamiltonian like the one we have
 assumed, any nearby band Hamiltonian with the same symmetries
 leads  qualitatively to the picture of fig. \ref{symbands}.   But this result is not forced by the universality class; a large
 enough deformation preserving the discrete symmetries will give an ordinary metal.   We show in fig. \ref{symbands3} how to modify the band
 structure, preserving its symmetry, so that the level crossings are no longer at the Fermi energy.

\subsection{Including Spin}

On contemplating the statement ``the band crossings will occur at $\veps_F$ if precisely one electron per unit cell is filling these two bands,'' one may wonder
if spin is being included in this counting.   Actually, our discussion has been so general that it makes
 sense with or without spin.   But there are two somewhat different cases.
 
In one case, spin-orbit couplings are important.   It is not a good approximation to consider spin to be decoupled from orbital motion. 
 The bands we have been drawing are the exact bands, taking spin and spin-dependent forces into account.  In the second case, spin-dependent forces are small and to begin with one ignores them and considers orbital motion only.
   In such a case, the two bands described by a Dirac Hamiltonian 
  are orbital bands.   When we include spin, in first approximation we simply double the picture, so that now there are four bands -- two copies
  of the familiar picture.    In this approximation, we get 2 chiral Weyl fermions, and they have the same chirality because (if the spin is
  decoupled from orbital motion) the spin up and spin down electrons have the same band Hamiltonian and so the same chirality. 

However, there always are spin-orbit forces in nature and generically the two pairs of bands will be split.    The exact problem is a
  four-band problem.   Assuming the density of electrons is such that 2 of the 4 bands are supposed to be filled, the crossings
  we care about (as they may be at or very near the Fermi energy) are those between the second and third bands, in order of increasing energy. 
The $N$ band version of the Nielsen-Ninomiya theorem that we come to in a moment  ensures that there will still be two
  Weyl crossings between the second and third bands (with the same chirality as before) but generically
  at slightly different energies and momenta.  The Fermi energy cannot equal the energy
  of each of these crossings, and generically it does not equal either of them, but  it will be close, assuming that the spin-orbit forces are weak.
  
   \begin{figure}
 \begin{center}
   \includegraphics[width=3.5in]{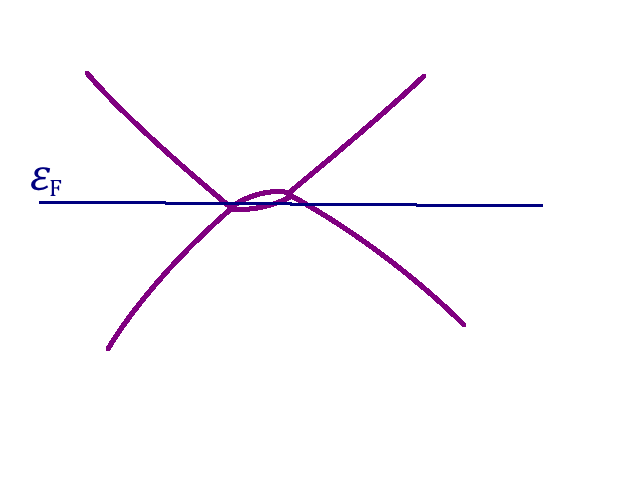}
 \end{center}
\caption{\small Two nearby band crossings that are not quite at the fermi energy. }
 \label{crude}
\end{figure}

 A very crude picture of two nearby Weyl crossings neither of which is  quite at the Fermi energy is in fig. \ref{crude}.
   Naively this leads to a normal metal with a very small density of charge carriers,
   but this is not the full story because Fermi liquid theory does not work well when the density of charge carriers is very small.
   
  \subsection{A System With Many Bands}
   
   Now let us discuss the generalization of the Nielsen-Ninomiya theorem for an $N$ band system.  We assume that the density of electrons
   is such that $k$ bands should be filled, for some integer $k<N$.    We let $\H_p$ be the full $N$-dimensional space of states at momentum $p$.
    At any value of $p$ such that the $k^{th}$ band (in order of increasing energy) does not meet the $k+1^{th}$, $\H_p$ has a well-defined subspace
   $\H'_p$ spanned by the $k$ lowest states.    The definition of $\H'_p$ does not make sense at points at which the $k^{th}$ band meets the
   $k+1^{th}$.     Just as before, to make this happen we have to adjust three parameters, so as in fig. \ref{badp}, 
   there will be finitely many bad points  in the Brillouin
   zone at which $\H'_p$
   is not defined.

Wherever $\H'_p$ is well-defined, it defines a $k$-dimensional subspace of $\H_p\cong \C^N$.    The space of all $k$-dimensional
   subspaces of $\C^N$ is called the Grassmannian $\Gr(k,N)$.   If $p_\alpha$ is an isolated  point on which $\H'_p$ is not defined,
   then $\H'_p$ is defined on a small sphere $S_\alpha$ around $p_\alpha$.
   Because 
   $\pi_2(\Gr(k,N))\cong \Z,$
   we can attach an integer-valued winding number $w(S_\alpha)$ to each $p_\alpha$. 
   
   Any of the explanations that we gave before for the case of two bands can be adapted to prove the Nielsen-Ninomiya theorem
   \begin{equation}\label{npn}\sum_\alpha w_\alpha=0.\end{equation}
   For example, let us consider the explanation based on the Berry connection.     Letting $\B'$ be as before the ``good part'' of the Brillouin zone
   with small neighborhoods of bad points removed, we have a rank $k$ complex vector bundle $\H'\to 
   \B'$ whose fiber at $p\in 
   \B'$ is $\H'_p$.  
   This is just the bundle spanned by the $k$ lowest bands.    On this bundle, there is a Berry connection, which is now a $U(k)$ gauge field.

      It is defined as follows.  To parallel transport $\psi_p\in \H'_p$ along a path $\gamma \subset B'$ (fig. \ref{berry}),  we require that\footnote{The condition
      that $\psi_p(s)$ should be in $\H'_{p(s)}$ for all $s$ determines the $s$ dependence of $\psi_p(s)$ up to the freedom to add an $s$-dependent
      element of $\H'_{p(s)}$. This freedom is fixed by eqn. (\ref{zomo}).}
   \be\label{zomo}\langle \psi'|\frac{\d}{\d s}|\psi\rangle=0,~~~{\mathrm{for~all}}~~\psi'\in\H'_p.\ee 
   In other words, $\d\psi/\d s$ is required to be orthogonal to $\H'_p$, for all $s$.   This gives a connection or $U(k)$ gauge field $\A$ on $\H'\to B'$. 
  It has a curvature $\F=\d\A+\A\wedge \A$.    The winding number $w(S_\alpha)$ is
  $$w(S_\alpha)=\int_{S_\alpha}c_1(\H')=\int_{S_\alpha} \,\frac{\Tr\, \F}{2\pi}.$$  
   
   Using the Bianchi identity $\d\Tr\,\F=0$, we get, with the help of Stokes's theorem
  \be\label{mpf}0=\int_{B'}\d\frac{\Tr\,\F}{2\pi}=\sum_\alpha\int_{S_\alpha} \frac{\Tr\,\F}{2\pi}=\sum_\alpha w(S_\alpha).\ee
  Thus, the proof using the Berry connection is the same as it was for two bands, except that we have to put a trace everywhere.
  
\begin{figure}
 \begin{center}
   \includegraphics[width=2.5in]{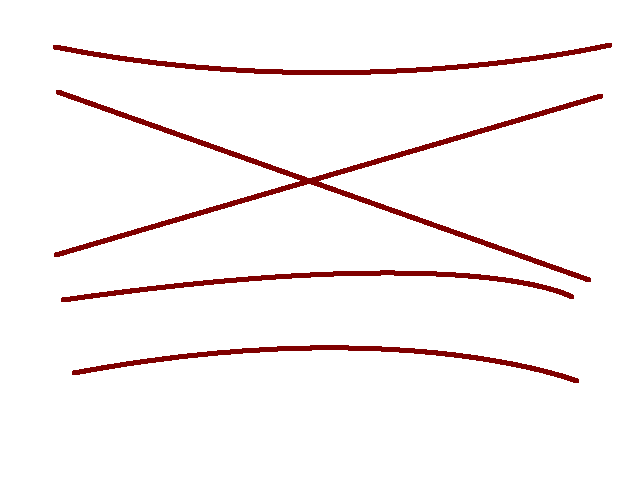}
 \end{center} 
\caption{\small A generic crossing between the $k^{th}$ band and the $k+1^{th}$ band, for any $k$, is governed by the same chiral Dirac Hamiltonian
that we originally encountered in the case of a two band system.  }
 \label{generi}
\end{figure}

  Generically, the winding number at a bad point is $\pm 1$, just as in the two band case.    The generic behavior at a crossing  of winding number
  $\pm 1$ is the familiar Weyl crossing between the $k^{th}$ and $k+1^{th}$ bands (fig. \ref{generi}).      So the points with winding number $\pm 1$
  give chiral Weyl fermions, and the Nielsen-Ninomiya theorem says that there are equally many of these of positive or negative chirality.
  
\subsection{Two Dimensions}\label{td}

None of this relied on discrete symmetries, though much of it becomes richer if one does consider materials with discrete
symmetries.   But what if we want to get massless Dirac fermions in 2 space dimensions rather than 3?  This will not work without discrete
symmetries because generically there would be no band crossings as we vary the 2 parameters of a 2-dimensional Brillouin zone.

In $2+1$ dimensions, there are only three $\gamma$ matrices $\gamma^0,\gamma^1,\gamma^2$, and they can be given a 2-dimensional representation.   So a
Dirac fermion in $2+1$ dimensions has only 2 components and the massless Dirac Hamiltonian  is
\be\label{twod}H=\sigma_1 p_1+\sigma_2 p_2.\ee    (To derive this from the relativistic Dirac equation $\gamma^\mu\partial_\mu\psi=0$, one basically
follows the derivation of eqn. (\ref{hameq}) in 3 space dimensions.) 
The energy levels are $\pm |p|$, and there is a level crossing at $p=0$.    We know that such a level crossing is nongeneric in 2 space dimensions, 
and concretely it
is possible to perturb the Dirac Hamiltonian by adding a mass term: 
\be\label{hamt}H=\sigma_1 p_1+\sigma_2 p_2+\sigma_3 m.\ee
The massive Dirac Hamiltonian has nondegenerate energy levels $\pm\sqrt{p^2+m^2}$.

However, the mass term violates some symmetries.   The {\it reflection} symmetry of
$H=\sigma_1 p_1+\sigma_2 p_2$
is 
\be\label{zono}\sR\psi(t,x_1,x_2)= \sigma_2\psi(t,-x_1,x_2)\ee
and the mass term 
\be\label{ozono}H'=m\sigma_3\ee
is odd under this.  The mass term is similarly odd under time-reversal.  With the Hamiltonian (\ref{hamt}) and the standard representation\footnote{In studying a $\sT$-invariant
theory, it is often more convenient to start with real $2\times 2$ gamma matrices $\g_\mu$, and that is what we will generally do.  In that case, the $\sigma$-matrices appearing in the Hamiltonian
are the real matrices $\sigma_i=\g_0\g_i$, as in eqn. (\ref{sigmam}).   With such a convention, $\sT$ acts by $\sT\psi(t,\vec x)=\pm\g^0\psi(-t,\vec x)$.  Hidden in this
statement is the following.  Suppose that we expand the complex (Dirac) fermion field $\psi$ in terms of two hermitian (Majorana) fermion fields $\chi_1,\chi_2$,
via $\psi=\chi_1+\i\chi_2$.   Then $\chi_1$ and $\chi_2 $ transform with opposite signs under $\sT$: $\sT\chi_1(t,\vec x)=\pm \g^0\chi_1(-t,\vec x)$, $\sT\chi_2(t,\vec x)=\mp \g^0\chi_2(-t,\vec x)$.  Since $\sT$ is antiunitary, $\sT\i=-\i\sT$,
the opposite signs in the transformation of $\chi_1$ and $\chi_2$ ensures the simple transformation that we have claimed for $\psi$.  Analogous statements hold
later  (footnote \ref{memo}) when we describe the action of $\sT$ in 3 space dimensions.}
of the
$\sigma$-matrices, time-reversal is 
\be\label{zamt}\sT\psi(t,\vec x)=\pm\sigma_1\psi(-t,\vec x). \ee
 The sign is actually physically meaningful and this turns out to be important in the theory of topological superconductors, though we will not explore that
 subject in the present lectures.

The physical reason that a mass term
violates reflection symmetry $\sR$ and time-reversal symmetry $\sT$ is as follows.    If  $\psi$ is a two-component electron field in two dimensions, then at
any given value of the spatial momentum $\vec p$, one component
of $\psi$ is a creation operator
and one is an annihilation operator.   Hence $\psi$ describes for each value of $\vec p$ only a single  state of charge 1 (along with a corresponding hole or antiparticle
of charge $-1$).     If the $\psi$ particle is massive, we can study it in its rest frame and its one spin state will transform with
 spin $1/2$ or $-1/2$ under the rotation
group. (In 2 space dimensions, the rotation group  is just the abelian group
$SO(2)$ and  has 1-dimensional representations.)   Either choice of sign is odd under $\sR$ or $\sT$,
so the mass term must violate those symmetries.     By contrast, if $m=0$, the fermion cannot be brought to rest, and in 2 space dimensions, we cannot define
its spin.\footnote{In $D$ spacetime dimensions, the ``spin'' of a relativistic particle is always described by the transformation of the quantum state under
the ``little group,'' the subgroup of the Lorentz group $SO(1,D-1)$ that preserves its energy-momentum $D$-vector $p^\mu$.  For a massive
particle, the little group is $SO(D-1)$ and the spin is a representation of this group.  For a massless particle, the little group is an extension
of $SO(D-2)$ by a noncompact group of ``translations,'' and the ``spin'' is actually determined by a representation of $SO(D-2)$.  (The ``translation'' part
of the little group acts trivially in all conventional relativistic field theories.  For an attempt to construct a theory in which this
would not be the case, see \cite{ST}.)  For $D=3$, $SO(D-2)$ is trivial and there is no notion of the ``spin'' of a massless particle.  In this
explanation, we have ignored reflection and time-reversal symmetry.  A massless particle in $D=3$ does have a meaningful transformation under
the discrete spacetime symmetries, when these are present.}    So the $m=0$ theory can be $\sR$- and $\sT$-conserving.   

This tells us that in a 2d crystal, it should be possible to find gapless Dirac-like modes as long as the crystal has a suitable $\sR$ or $\sT$ symmetry,
and the gapless modes occur at an $\sR$- or $\sT$-invariant value of the momentum.    It is not hard to give examples.  The most famous example
is graphene; we will discuss this case in Lecture Three.  For now, 
   I will just remark that rather as for Weyl points in 3 space dimensions,
there are two versions, either a material that with spin included has an $\sR$ or $\sT$ symmetry that
leads to a gapless mode, or a material with small spin-orbit forces that has the appropriate
 property if spin and spin-orbit forces are ignored.  In the latter  case, in the real world, one will
get modes with a gap that is very small but not quite zero.  This is indeed what happens in graphene \cite{KM}.

\subsection{Weyl Fermions And Fermi Arcs}

\begin{figure}
 \begin{center}
   \includegraphics[width=2.5in]{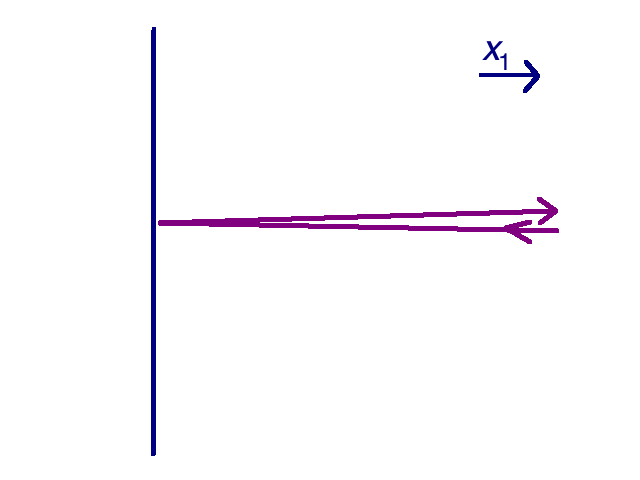}
 \end{center}
\caption{\small A particle reflecting at right angles from the boundary must reverse its helicity (the component of its angular momentum along the direction
of motion) if it is to conserve its angular momentum. }
 \label{RightAngles}
\end{figure}

Now we will begin our discussion of topologically-determined edge modes in condensed matter physics.
 We will do this in 3 space dimensions and we will start by considering
a non-chiral massless Dirac fermion $\psi$.   For now, never mind how to realize this in condensed matter physics.   We suppose that $\psi$ is
confined to a half-space (possibly the interior of a crystal) and we ask what kind of boundary condition it should obey when it is reflected from a boundary. 
  For reflection at right angles, as sketched in fig. \ref{RightAngles},
 a simple boundary condition would conserve angular momentum.  

 By ``conserving angular momentum,'' I mean that for a boundary condition at $x_1=0$, the component $J_1$ of angular momentum
 around the $x_1$ axis should be conserved.\footnote{In a crystal, $J_1$  might be conserved only mod $n$, for $n=2,3,4$, or $6$. The argument
 below will show that the boundary condition on a Weyl fermion cannot conserve $J_1$ mod $n$ for any $n>1$.}    Since the direction of motion 
 is reversed in the scattering, the helicity has to be reversed.   For a Dirac fermion, that is possible, because a Dirac fermion has both helicities.
See eqn. (\ref{molof}) below for the angular momentum conserving but helicity-violating boundary condition that is possible for a Dirac fermion.
  
 But what sort of boundary condition can we have for a massless Weyl fermion, which has only one helicity?    Obviously, the boundary
 condition cannot reverse the helicity, and therefore it cannot conserve angular momentum.  Any boundary condition will have to pick
 a preferred direction in the boundary plane.    For a chiral Dirac Hamiltonian
 \be\label{chid}H=-i\vec\sigma\cdot \frac{\partial}{\partial\vec x},\ee
 a good boundary condition at $x_1=0$ is
 \be\label{goodbc}\left.M\psi\right|_{x_1=0}=\left.\psi\right|_{x_1=0}\ee with
 \be\label{mis}M=\sigma_2 \cos\alpha+\sigma_3\sin\alpha\ee
for some angle $\alpha$.

What makes this a good boundary condition is that it makes $H=-i\vec\sigma\cdot\vec\nabla$ hermitian.   To prove that $H=-i\vec\sigma\cdot\vec\nabla$ is hermitian,
\be\label{zeff}\langle\psi_1,H\psi_2\rangle=\langle H\psi_1,\psi_2\rangle,\ee
one has to integrate by parts.  A potential boundary term in this integration by parts vanishes because
$\{M,\sigma_1\}=0$,
and our choice $M=\sigma_2 \cos\alpha+\sigma_3\sin\alpha$ was made to ensure this.    In particular, this will not  work if we pick $M=\sigma_1$,
and that again shows that the boundary condition cannot be invariant under rotation of the $x_2-x_3$ plane.   It cannot even preserve a nontrivial
discrete subgroup of this rotation symmetry, such as might be present in a crystal.

Hermiticity would let us add additional momentum-dependent terms to the operator $M$ that appears in the boundary condition, but in the low momentum
limit, near the band-crossing point, it forces $M$ to take the form that we have indicated, with some value of $\alpha$.
In continuum field theory, we could regard $\alpha$ as a free parameter.  That is not really the situation in condensed matter physics.  In
a concrete model whose band structure in bulk leads to the existence of a band-crossing point described by a chiral Dirac Hamiltonian,
solving the Schrodinger equation near the boundary of the system will determine an effective value of $\alpha$.   However, modifying the boundary,
for example by adding an extra layer of atoms on the surface of a material, would generically change that value.

The value of $\alpha$  can be absorbed in a rotation of the $x_2-x_3$ plane, so in analyzing the consequences of the boundary condition,
we can consider the special case $\alpha=0$, meaning that the boundary condition is $\sigma_2\psi\vert=\psi\vert$.
Something very interesting happens when we solve the Schrodinger equation with this boundary condition.     
Let us try to solve the equation $H\psi=0$ assuming that
$\sigma_2\psi=\psi$ everywhere (not only on the boundary) and also assuming that $\partial\psi/\partial x_2=0$.   
Then
\be\label{wolt}H\psi=-\i\left(\sigma_1\partial_1+\sigma_2\partial_2+\sigma_3\partial_3\right)\psi=-\i\sigma_1(\partial_1-i\partial_3)\psi.\ee 
(Recall that $\sigma_3=-i\sigma_1\sigma_2$, so $\sigma_3\psi=-i\sigma_1\psi$.) So we can solve $H\psi=0$ with
\be\label{polk}\psi=\exp(\i k_1 x_3+k_1x_1)\psi_0,\ee
where $\psi_0$ is a constant spinor 
obeying 
\be\label{folt}\sigma_2\psi_0=\psi_0.\ee
Moreover, this solution
is plane-wave normalizable if 
\be\label{formix}k_1<0.\ee
More generally, for any real $\veps$, we can solve $H\psi=\veps\psi$ with
\be\label{molt}\psi=\exp(\i\veps x_2)\exp(\i k_1 x_3+k_1x_1)\psi_0.\ee

Assuming that our system is supported in the half-space $x_1\geq 0$, with boundary at $x_1=0$,
these solutions decay exponentially
away from the boundary as long as $k_1< 0$.
They provide our first examples of topologically-determined edge-localized states in condensed matter physics.  

For each energy $\veps$ that is close enough to the band-crossing energy so that the above analysis is applicable, 
we have found  edge localized states that are supported on a ray $k_1< 0$, $k_2=\veps$ in the $k_1-k_2$ plane.
At the endpoint $k_1=0$ of the ray, edge-localization breaks down and the solution becomes a plane wave (with momentum in the $x_2$ direction
only).  As such it is
indistinguishable from a bulk state.

From the point of view of condensed matter physics, the fact that the edge-localized states of given energy lie on a staight line in momentum space
is certainly not universal.  We could add all sorts of higher order terms to the Hamiltonian and the boundary condition, and this would modify the dispersion
 relation
of the edge-localized states, just as it would modify the dispersion relation for bulk states.  However, the analysis that we have made is universal
near the band-crossing point at $\vec k=0$.  This analysis shows that at any energy sufficiently
near the band-crossing energy, there is an arc of edge-localized
states, known as a Fermi arc \cite{WTVS}.  

An arc parametrizing edge-localized states
can only end when the  state in question ceases to  be edge-localized.  But at that point, as in our example, the edge-localized
state  becomes indistinguishable from some bulk plane wave state.  In the presence of a boundary at $x_3=0$, the quantities $k_1$,
$k_2$, and $\veps$ are conserved  but $k_3$ is not.  So the values of $k_1$, $k_2$, and $\veps$ at the endpoint of a Fermi arc will coincide
with the corresponding values for some bulk plane wave state, with some value of $k_3$.  (In general, though not in our simple 
model, this value will depend on $\veps$.)

A Fermi arc that has an end associated to a band-crossing point will inevitably have a second end, which will be associated to some other band-crossing
point.   Let us examine this matter from the point of view of the Nielsen-Ninomiya theorem.  In general, we know that there are
always multiple Weyl points in the Brillouin zone,
say at momenta $\vec k_\alpha$, $\alpha=1,\dots,s$.  In the presence of a boundary at $x_1=0$, the perpendicular part $k^\perp=k_1$ 
of the momentum  is not conserved and we should classify the Weyl points only by $k^\parallel=(k_2,k_3)$.

\begin{figure}
 \begin{center}
   \includegraphics[width=2.5in]{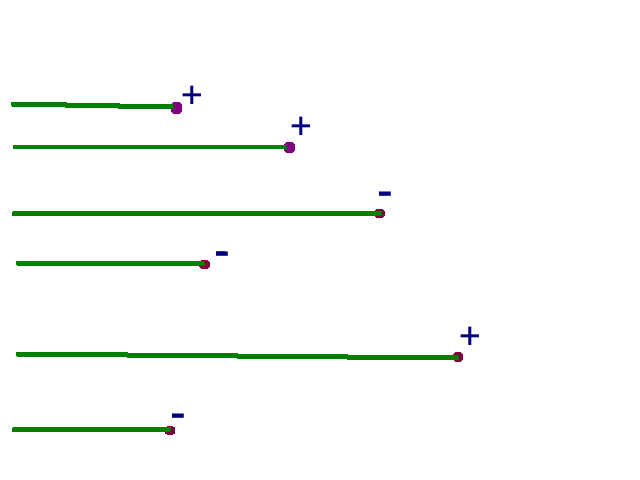}
 \end{center}
\caption{\small Projection of band-crossing points from the bulk Brillouin zone, which is parametrized by $\vec k$, to the surface Brillouin zone,
which is parametrized by $k^\parallel$ only. (In the picture, $k^\perp$ runs horizontally and $k^\parallel$ runs vertically.)
  Generically the band-crossing points occur at distinct values of $k^\parallel$.  In this case, their
projections are ends of Fermi arcs.}
 \label{projection}
\end{figure}

In bulk, because momentum is conserved, gapless modes at different values of $\vec k$ do not ``mix'' with each other and can be treated separately.
But when we consider the behavior near a boundary, the ``perpendicular'' component $k^\perp$ of the momentum is not conserved and we should
only use $k^\parallel$.  So we  project the bad points to 2 dimensions, as in fig. \ref{projection}.  As long as the projections $k_\alpha^\parallel$ of the Weyl points are all distinct, they
will be connected pairwise by Fermi arcs. 

\begin{figure}
 \begin{center}
   \includegraphics[width=3.5in]{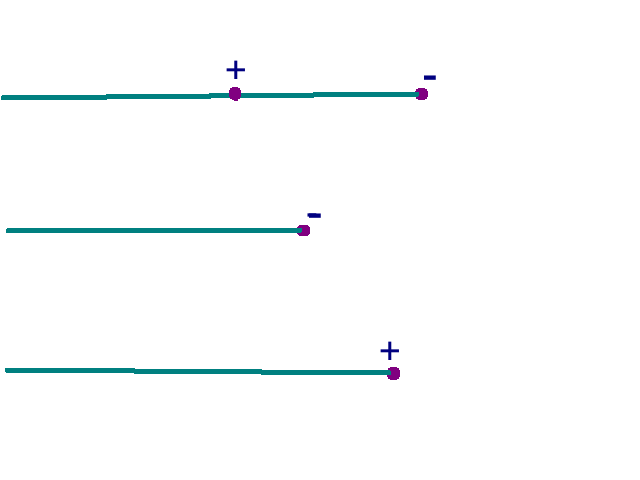}
 \end{center}
\caption{\small Discrete symmetries may make it natural for two or more band crossing points to occur at the same value of $k^\parallel$. }
 \label{special}
\end{figure}

 But if two Weyl points of opposite chirality project to the same point in the boundary
momentum space, as in fig. \ref{special}, 
then there is no need for either one to connect to a Fermi arc.    From a low energy point of view, the two modes of opposite
chirality combine to a Dirac fermion with both chiralities.  A Dirac fermion admits a rotation-invariant boundary condition and can be gapped,
as we discuss in section \ref{df}.

Of course, whether two given Weyl points coincide when projected to the boundary depends on which boundary
face we consider.    But the discrete symmetries that make Weyl points interesting can also make it natural, for some crystal facets, that
two Weyl points have the same projection.   In fact, this will have to happen if a crystal has a nontrivial group of rotations that preserves the plane
$x_1=0$.  Since a single Weyl fermion would not admit a boundary condition that preserves $J_1$ mod $n$ for any $n>1$, in the presence
of such a conservation law, we will never get just one Weyl fermion at a given value of $k^\parallel$.

\subsection{Gapless Boundary Modes From Dirac Fermions}\label{df}

It is also possible to get boundary-localized modes from Dirac fermions, and  this is important in understanding topological insulators.    
In the absence of discrete symmetries, and without tuning any parameters,
it is not natural in condensed matter physics to get a massless Dirac (as opposed to Weyl) fermion in three space dimensions.    For a four-component
fermion field $\psi$ with both chiralities, mass terms are possible; in fact there are two such terms.  The general Lorentz-invariant massive
Dirac equation is
\be\label{zolof}\left(\sum_\mu\gamma^\mu\partial_\mu -m-\i m'\gamma_5\right)\psi=0.\ee
The corresponding Hamiltonian is
\be\label{olof}H=\g^0\vec\gamma\cdot \vec p-\i m\g^0-\i m'\g_1\g_2\g_3. \ee
 Relativistically, to get a massless Dirac fermion, we need a reason for $m=m'=0$.  Symmetries (spacetime or chiral symmetries)
 are the obvious place to look.

Let us consider the case of assuming time-reversal symmetry $\sT$. This is enough to set one of the two parameters to 0 but not both.
 The Dirac equation becomes\footnote{\label{memo}
 This is invariant under $\sT\psi(t,\vec x)=\g^1\g^2\g^3\psi(-t,\vec x)$. 
  We assume that the gamma matrices are real $4\times 4$ matrices obeying $\{\g_\mu,\g_\nu\}=2\eta_{\mu\nu}$, where $\eta_{\mu\nu}=\mathrm{diag}(-1,1,1,1)$.}
\be\label{lookme}\left(\sum_\mu\gamma^\mu\partial_\mu -m\right)\psi=0.\ee
  Generically $m$ is not 0 but of course if we adjust one parameter, we can make $m$ vanish.  
  
    In condensed matter physics, what we adjust to make a parameter in the Hamiltonian vanish might be, for example, the chemical composition
  of an alloy.  However, we observe that while eqn. (\ref{olof}) is the general Lorentz-invariant Hamiltonian for this system, in condensed matter physics we should not assume Lorentz-invariance of the Hamiltonian and more terms are possible.  
  The analysis of a more general Hamiltonian is more complicated.  But there is a useful lesson that we can learn from a further study of the relativistic case.
  
Consider a Dirac fermion confined to the half-space $x_1\geq 0$. 
The Dirac operator  admits a natural, rotation-invariant boundary
condition
\be\label{molof}\g_1\psi|_{x_1=0}=\pm \psi|_{x_1=0},\ee
with some choice of sign.   This stands in contrast to a Weyl fermion, which as we discussed earlier does not admit a rotation-symmetric
boundary condition.  The boundary condition (\ref{molof}) is helicity-violating, because $\g_1$ anticommutes with the chirality or helicity operator
$\g_5=\i\g^0\g^1\g^2\g^3$.  So it would not make sense for a Weyl fermion, which is an eigenstate of $\g_5$ and has only one helicity.  Note that the boundary 
condition (\ref{molof}) is $\sT$-conserving, with the action of $\sT$ defined in footnote \ref{memo}, because $\g_1$ commutes with $\g^1\g^2\g^3$.  
The sign in the boundary condition can be reversed by $\psi\to\g_5 \psi$, which also changes the sign of the mass parameter $m$ in eqn. (\ref{lookme}).
So as long as we consider both signs of $m$, we can choose a $+$ sign in the boundary condition.

Let  $x_\parallel $ be the coordinates along the boundary and
\be\label{zogg}\gamma\cdot\partial^\parallel=\sum_{\mu\not=1}\gamma^\mu\partial_\mu\ee
the $2+1$-dimensional Dirac operator along the boundary. 
We can obey the $3+1$-dimensional Dirac equation in the half-space $x_1\geq 0$ with
\be\label{ogg}\psi=\exp(mx_1)\psi_\parallel(x_\parallel)\ee
where 
\be\label{pogg}\gamma_1\psi_\parallel=\psi_\parallel,~~~\gamma\cdot\partial^\parallel\psi_\parallel(x_\parallel)=0.\ee
For $m>0$, this solution is highly unnormalizable.   But for $m<0$, it is plane-wave normalizable and localized along the boundary.

To be more precise, since $\psi_{\parallel}$ was constrained to obey the massless $2+1$-dimensional Dirac equation
\be\label{plogg}\gamma\cdot\partial^\parallel\psi_\parallel(x_\parallel)=0,\ee
we get a $2+1$-dimensional massless Dirac fermion.   (This is a standard 2-component Dirac fermion in $2+1$ dimensions, because half of the components 
of the original 4-component $3+1$-dimensional Dirac fermion are removed by the
 constraint $\gamma_1\psi_{\parallel}=\psi_{\parallel}$.) 

In a $\sT$-invariant theory, a phase that is gapped in bulk and has a single boundary-localized massless Dirac fermion is essentially
different from a phase that is gapped both on the bulk and on the boundary.  That is because once we find a single boundary-localized massless
fermion, small $\sT$-invariant perturbations, even if they violate Lorentz symmetry, will
not cause the boundary to be gapped.  Indeed, we recall that $\sT$-invariance does not allow a single $2+1$-dimensional 
Dirac fermion to acquire a mass.

  However, $\sT$-invariance
would permit a {\it pair}  of Dirac fermions in $2+1$ dimensions to acquire bare masses.  The $\sT$-invariant Dirac equation for such a pair is\footnote{$\sT$
acts by $\sT\psi_i(t,\vec x)= \g^0\psi_i(-t,\vec x)$, $i=1,2$.  The mass term has been chosen to be hermitian while ensuring $\sT$-invariance.}
\begin{equation}\label{zuffo}\left(\sum_{\mu=0}^2\g^\mu\partial_\mu-\i  \bp0 &m\cr -m& 0\ep \right)\bp\psi_1\cr\psi_2\ep=0.\ee
(Upon diagonalizing the mass term, one learns that this gives two $2+1$-dimensional massive Dirac fermions with equal and opposite masses.)
Thus, in a condensed matter system of 3 space dimensions with $\sT$-invariance and no other special properties, the number of boundary-localized
Dirac fermions will be generically either 0 or 1.   These are two different phases; one cannot pass between them, maintaining $\sT$-invariance,
as long as 
the bulk theory is gapped.  (When the bulk is gapless, a boundary gapless mode can disappear by becoming indistinguishable from a bulk 
mode.)
The case that there is a gapless Dirac mode on the boundary is the 3-dimensional topological insulator
\cite{FMK}.   Topological band theory gives a powerful way to understand this phase, but we will not explore that here.

What we have said shows  that when the mass of a bulk Dirac fermion passes through 0, this results in a phase transition between
an ordinary insulator and a topological insulator.  That is a particularly simple path between these two phases that looks natural
from a relativistic point of view.
But this path is nongeneric in the context of condensed matter physics, provided $\sT$ is the only pertinent symmetry,
because in condensed matter physics the Hamiltonian
will generically contain additional terms that do not respect Lorentz invariance.  It turns out \cite{Mur,KK} that generically the transition
from an ordinary insulator to a topological one is a more complicated process with an intermediate conducting phase.\footnote{\label{complicated}
Starting with
a trivial insulator, as one varies
a parameter, the system undergoes a transition described in section \ref{somex}, with appearance of a pair of Weyl points of opposite chirality.
This happens simultaneously at two equal and opposite values of $\vec p$, exchanged by $\sT$.  Going further, the Weyl points reconnect
and annihilate, leaving a topological insulator.}  With $\sP$ assumed as well as $\sT$, the simple model in which the phase transition between
the two types of insulator involves a massless Dirac fermion is indeed valid.  We return to this point at the end of section \ref{dlat}.

  Generically, in the context of condensed matter physics, 
the fermi energy $\veps_F$ of a topological insulator does not pass through the Dirac point in the boundary theory.
So (fig. \ref{specialo}) the boundary of a topological insulator is more like an ordinary metal than
the Weyl semimetals that we talked about before.   Relativistically, the boundary theory is analogous to the theory
of a massless Dirac fermion with a nonzero chemical potential.

  \begin{figure}
 \begin{center}
   \includegraphics[width=2.5in]{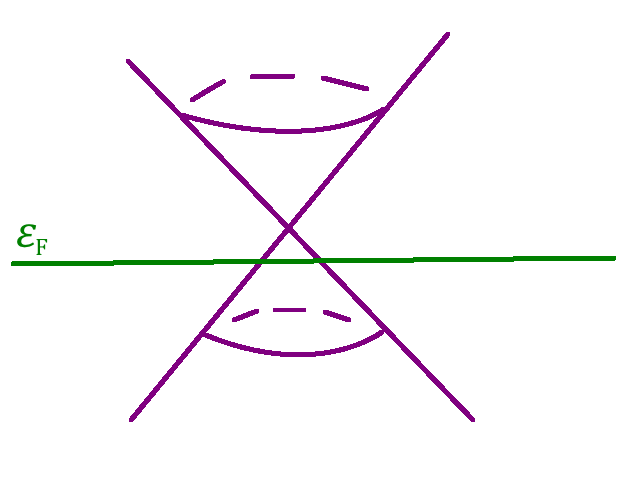}
 \end{center}
\caption{\small Generically, the band crossing point of the edge-localized mode of a topological insulator in 3 space dimensions does not
occur at the Fermi energy.  Accordingly, the boundary has much in common with a normal metal.}
 \label{specialo}
\end{figure}

\subsection{Discrete Lattice  Symmetries and Massless Dirac Fermions}\label{dlat}

Because the Hamiltonian need not be Lorentz-invariant,  we have really not yet  found a way to generate massless Dirac fermions
in condensed matter physics, even assuming time-reversal symmetry and varying one parameter.
 However, massless Dirac fermions can become natural if more
symmetry is assumed.  In what follows, we sketch a construction in \cite{YN}, leaving some details to the reader.    For some of the
background, see  \cite{ASSS}.

First, let us reconsider the relativistic massless Dirac fermion in $3+1$ dimensions.  The Hamiltonian is $H=\g_0\vec \gamma\cdot \vec p$, where now
the $\g^\mu$ are $4\times 4$ matrices and we assume no chiral projection.  Equivalently, the Hamiltonian is conjugate to
\be\label{perf}H=\bp \vec\sigma\cdot \vec p & 0 \cr 0 & -\vec\sigma\cdot\vec p\ep.\ee
The energy levels are $\pm|\vec p|$, each occuring with multiplicity 2.

One explanation of the twofold degeneracy of the bands is the following.   The massless nonchiral Dirac Hamiltonian has both time-reversal
symmetry $\sT$ and parity symmetry $\sP$.  Each of these reverses the sign of the spatial momentum, so the product $\sP\sT$ is a symmetry
at any given momentum.  But $\sP\sT$ is an antiunitary symmetry, and in acting on a fermion state, $(\sP\sT)^2=-1$.  The existence of an
antiunitary symmetry that leaves $\vec p$ fixed and squares to $-1$ means that the energy levels at each momentum have a Kramers degeneracy, which accounts
for the doubling.

It is perfectly natural in condensed matter physics to consider a material with $\sPT$ symmetry.  Indeed, many nonmagnetic materials
have $\sT$ symmetry, and many crystals have $\sP$ symmetry (usually called inversion symmetry in the context of condensed matter).  
In a $\sPT$-symmetric material, all bands will have 2-fold degeneracy.  If spin-orbit forces can be ignored, this is simply the 2-fold
degeneracy resulting from spin.  However, $\sPT$ symmetry forces an exact 2-fold degeneracy of all bands even with spin-dependent
forces included.

The question arises of whether  in a $\sPT$-symmetric material, and for the moment assuming no further symmetry,
 the generic 2-fold degeneracy of the bands becomes a 4-fold
degeneracy somewhere in the Brillouin zone.  To answer this question, we consider a generic $\sPT$ symmetric system with 4 bands.
The unitary group acting on 4 states (that is, on the states of 4 bands at a given value of $\vec p$) is in general $U(4)$.  The subgroup
of this group that commutes with an antiunitary symmetry $\sPT$ that satisfies $(\sPT)^2=-1$ is $Sp(4)$.  The fifteen $4\times 4$ traceless
hermitian matrices that might conceivably lift the band degeneracy transform under $Sp(4)$ as ${\mathbf {15}}={\mathbf{5}}\oplus
{\mathbf{10}}$. The $\sPT$-invariant ones transform as ${\mathbf{5}}$.  In fact, the group $Sp(4)$ is the double
cover $\mathrm{Spin}(5)$ of $SO(5)$, and the ${\mathbf{5}}$ of $Sp(4)$ is just the defining 5-dimensional representation of $SO(5)$.

For a basis of  5 traceless, hermitian, $\sPT$-invariant  $4\times 4$ matrices, we can take gamma matrices\footnote{We place hats
on these gamma matrices as they do not coincide with the  $SO(3,1)$ gamma matrices  that were used in the relativistic
description.  We explain the relationship in eqn. (\ref{preet}).} $\h\g_1,\dots,\h\g_5$,
obeying $\{\h\g_i,\h\g_j\}=2\delta_{ij}$.  The general $\sPT$-invariant traceless 4-band Hamiltonian is therefore
\be\label{gert}H=\sum_{i=1}^5A_i(\vec p)\h\g_i.  \ee
To get a 4-fold band degeneracy, we need to make the five functions $A_i(\vec p)$ simultaneously vanish.  Generically, in spatial dimension 3,
this
will not happen anywhere in the Brillouin zone.  Therefore, to get a massless Dirac fermion, we need to assume more symmetry.

Let us consider a crystal with a $\Z_n$ rotation symmetry, where the most convenient values\footnote{A similar story holds for $n=3$ with
slight modifications.} of $n$ are 4 and 6.  We assume that
the band degeneracy of interest will occur at a $\Z_n$-invariant value of the momentum.  (In general, only the subgroup of $\Z_n$ that
leaves fixed the momentum at which a given band degeneracy occurs will be relevant in protecting that degeneracy.)  We can
arrange the crystal axes so that $\Z_n$ leaves $p_1$ fixed.  We can also assume, without essential loss of generality, that the $\Z_n$-invariant
value of the momentum at which there will be a massless fermion satisfies $p_2=p_3=0$.  Further, we can assume that near $p_2=p_3=0$,
$\Z_n$ is generated by an element $R$ that acts as a $2\pi/n$ rotation of the $p_2-p_3$ plane, leaving $p_1$ fixed.

$R$ will be realized on the 4 fermion bands as an element of $Sp(4)$; it will act on the 5 gamma matrices as an element of $SO(5)$.
A general element of $SO(5)$ of order $n$  has eigenvalues $1, $ $\exp(\pm 2\pi \i r/n)$, $\exp(\pm 2\pi \i s/n)$ in the ${\mathbf 5}$
representation, for some integers $r,s$ which we can take to be nonnegative and\footnote{We can further restrict to $r\leq n/2$.
We could restrict $r,s$ to be both  $\leq n/2$, but then
we would need to include a possible overall minus sign in eqn. (\ref{mulch}) below, because of the fact that the group $Sp(4)$ that acts on the bands
is  a double cover of the group $SO(5)$ that acts on the gamma matrices.} $\leq n$. For a reason that we will explain in a moment, we can
assume that $r+s$ is odd.   Different cases are of interest for condensed matter
physics, but to get a massless Dirac fermion we need $r$ (or $s$) to equal 1 while the other is nonzero. Since $r+s$ is odd, if $r=1$ and $s\not=0$
then $s$ is even and $\geq 2$.  

The reason that $r+s$ should be odd is as follows.  If the gamma matrices transform under $R$ as $1,$ $\exp(\pm 2\pi \i r/n)$,
$\exp(\pm 2\pi\i s/n)$, then the 4 bands that they act on transform under $R$ as $\exp(\pi \i(\pm r\pm s)/n)$.  But on fermions one wants $R^n=-1$, which
corresponds to $r+s$ odd.  For $r=1$, the eigenvalues of $R$ acting on the 4 bands are
\be\label{mulch}\exp(\pm\pi \i/n)\exp(\pm 2\pi\i s'/n),~~~s'=s/2,\ee
where $s'$ is an integer since $s$ is even.

We can pick a basis of the 5 gamma matrices so that $\h\gamma_1$ is $R$-invariant, the $\h\g_2-\h\g_3$ plane is rotated by $R$ by an angle
$2\pi/n$, and the $\h\g_4-\h\g_5$ plane is rotated by $R$ by an angle $2\pi s/n$.  Let us now analyze the Hamiltonian, first along the axis
$p_2=p_3=0$, and then slightly away from this axis.  

At $p_2=p_3=0$, the momentum is $R$-invariant.  The only $R$-invariant gamma matrix is $\h\g_1$, so a general $R$-invariant Hamiltonian is $H=\h\g_1 A(p_1)$.  It is natural
for $A(p_1)$ to have a simple zero at some value of $p_1$, and that is where a massless Dirac fermion will occur.  Expanding the Hamiltonian
in powers of $p_2$ and $p_3$, because of the assumption that $r=1$, the $\h\g_2-\h\g_3$ plane is rotated by $R$ just like the $p_2-p_3$
plane and the Hamiltonian can have a term $B(p_1)(\h\g_2p_2+\h\g_3p_3)$.  The coefficients of $\h\g_4,\h\g_5$ vanish up to higher order in $p_2,p_3$
because $s\geq 2$.  Thus to linear order in $p_2,p_3$, the Hamiltonian is
\be\label{wonder}H=A(p_1)\h\g_1+B(p_1)(\h\g_2p_2+\h\g_3p_3).\ee
There is a massless Dirac fermion at any simple zero of $A(p_1)$, assuming that $B(p_1)\not=0$ at that point.  

The assumption that $r$ (or $s$) equals 1 is actually rather natural, for the following reason.  It led to the result (\ref{mulch})
for the transformation of the 4 bands under $R$.  But this is the result that one would expect if the 4 bands of interest are the tensor
product of 2 spatial bands with 2 spin states. The  eigenvalues of a $2\pi/n$ rotation acting on the spin states of a spin 1/2
particle are $\exp(\pm \pi\i/n)$, and the eigenvalues of such a rotation acting on 2 spatial bands in a $\sPT$-invariant system will
be $\exp(\pm 2\pi \i s'/n)$ for some integer $s'$.

From a relativistic point of view, one would account for what we have found by saying that what
microscopically is the spatial rotation $R$ has behaved in the effective theory at low energies as a combination of a spatial rotation
and a chiral symmetry.  The chiral symmetry, which is expressed in the above analysis as the rotation of the $\h\g_4-\h\g_5$ plane,
arises if $s\not=0$.   To explain this point more fully, observe that the relation between the five terms in the $\sPT$-invariant
relativistic Dirac Hamiltonian (\ref{olof}) and the five terms in eqn. (\ref{gert}) comes from 
\be\label{preet}\g_0\g_i=\h\g_i~ ~(i=1,2,3), ~-\i\g^0=\h\g_4,~\i\g_1\g_2\g_3=\h\g_5.\ee
Thus the rotation of the $\h\g^4-\h\g^5$ plane is a rotation of the parameters $m$ and $m'$ in the relativistic Hamiltonian (\ref{olof}).  In relativistic physics,
a symmetry
that rotates those parameters is usually called a chiral symmetry and such a symmetry forces the vanishing of the fermion mass.

In the preceding analysis,
we have made use only of $\sPT$ and not of separate $\sP$ and $\sT$ symmetry.   This is appropriate if a material
has only $\sPT$ symmetry and not separate $\sP$ and $\sT$ symmetries, or if the material has both symmetries but the band
degeneracy of interest occurs at a value of the momentum that is $\sPT$-invariant but not $\sP$- or $\sT$-invariant.  A further
interesting construction \cite{YZ} becomes possible in a material that does have separate $\sP$ and $\sT$ symmetry, and governs
band degeneracies that occur at values of the momentum that are invariant under both symmetries.  It turns out to be necessary
to consider non-symmorphic symmetries (symmetries that mix rotations and partial lattice translations in an essential way).
For an introduction, focusing on an analogous problem in 2 space dimensions, see \cite{YK}.

As a simple example of the consequences of assuming both $\sP$ and $\sT$ symmetry, we reconsider the transition, discussed in
section \ref{df}, between
an ordinary insulator and a topological one.  In fact, with both $\sP$ and $\sT$ symmetry,\footnote{We assume that the bands are not all even or all odd under $\sP$, so that
$\sP\not=\pm 1$ as an element of $Sp(4)$.  Otherwise the low energy physics is quite different.} this phase transition occurs
when a Dirac fermion mass passes through 0, rather than by the more complicated route described in footnote \ref{complicated}.
The point is that $\sPT$ symmetry enables us to express the Hamiltonian in terms of five functions,
as in eqn. (\ref{gert}) (with eqn. (\ref{preet}) as a recipe to express the relativistic Hamiltonian (\ref{olof})  in terms of the five functions $A_i$ of eqn. (\ref{gert})).  When we assume
separate $\sT$ and $\sP$ symmetry, 
only one\footnote{A simple way to see this is to observe that $\sP$ must be an element of $Sp(4)$ obeying $\sP^2=1$, $\sP\not=\pm 1$.    Any such element (other than $\sP=\pm 1$)  acts on the
gamma matrices as
 $\mathrm{diag}(-1,-1,-1,-1,1)$  (up to conjugation) and so leaves invariant precisely one linear combination of the $A_i$.  With standard relativistic conventions that
 were used in writing eqn. (\ref{olof}), $\sP$
 acts by $\sP\psi(t,\vec x)=\i\g^0\psi(t,-\vec x)$ and the $\sP$-invariant coupling is $A_4=m$.}
 of the five functions $A_i$, namely $A_4=m$, is $\sT$- and $\sP$-invariant.  The other $A_i$ are all odd and
vanish at the $\sT$- and $\sP$-invariant point $\vec p=0$.  So at  $\vec p=0$,
 a 4-fold band degeneracy can be achieved by setting to 0 the one parameter $m$.

\subsection{Simple Examples Of Band Hamiltonians}

We conclude by giving simple examples of band Hamiltonians to illustrate some of these ideas.

One goal is to describe a simple band Hamiltonian that can be approximated near $\vec p=0$ by the chiral Dirac Hamiltonian
$H=\vec\sigma\cdot \vec p$.  For this purpose, the main difference between band theory 
 and the relativistic problem is that in band theory the components of the momentum are periodic variables.
We assume a simple cubic lattice of lattice spacing $a$ so that the linear components of the momentum have period $2\pi/a$.
To write a band  Hamiltonian, we can replace the linear component $p_i$, $i=1,2,3$ of the electron momentum
by $\frac{1}{a}\sin (p_ia)$, which has the correct periodicity and is equivalent to $p_i$ for small $p_i$.  This motivates
the band Hamiltonian
\be\label{twof}H=\frac{1}{a}\sum_{i=1}^3 \sigma_i\sin(p_ia).  \ee
The formula $\sin u=(e^{\i u}-e^{-\i u})/2\i$ shows that, when written in coordinate space, this Hamiltonian describes
nearest neighbor hopping on the cubic lattice.  

For $\vec p\to 0$, this Hamiltonian can be approximated by $\vec\sigma\cdot \vec p$, so there is a Weyl point of
positive chirality at $\vec p=0$.  However, a little reflection shows that the model actually has a total of 8 Weyl points in the Brillouin
zone; they are the 8 points at which each of the $p_i$ equals 0 or $\pi/a$.    Using the criterion (\ref{perd}), the reader can verify that
4 of these Weyl points have positive chirality and 4 have negative chirality.  Thus the net chirality is 0, in keeping with the Nielsen-Ninomiya
theorem.  Indeed, the Nielsen-Ninomiya theorem was inspired by  examples such as this one. 

We can similarly write a periodic version of the non-chiral 4-component Dirac Hamiltonian (\ref{olof}):
\be\label{ollof}H=\frac{1}{a}\sum_{i=1}^3\g_0\gamma_i\sin(p_ia) -\i m\g^0. \ee
We have set $m'=0$ to ensure $\sT$ and $\sP$ invariance (assuming that $m$ is an even function of $\vec p$).  If $m=0$, there are massless
Dirac fermions at the same 8 points as before.  If $m$ is nonzero, the system is gapped for all $\vec p$.  It is an insulator and in fact a trivial one
if $m$ is a constant.  However, we get something new if $m$ is a more general periodic function of $\vec p$.  Assuming that $m$ is small, the term in $H$
proportional to $m$ is only important near the 8 points  that support an almost massless Dirac fermion.  Moreover, all that
really matters is the sign of $m$ at those 8 points.  We would like $m(\vec p)$ to have a finite Fourier expansion in powers of $\exp(\pm \i p_ia)$
(so that the position space Hamiltonian has finite range).  Even with this constraint, there is no problem to vary independently the sign of $m(\vec p)$
at the 8 points of interest.  When one of those signs passes through 0, an edge-localized massless Dirac fermion appears or disappears.  Thus
this  Hamiltonian with a suitable function $m(\vec p)$ gives a simple  model of a topological insulator in 3 space dimensions.  

What we have described is somewhat analogous to the Haldane model  \cite{Haldane} of a topologically non-trivial band insulator in 2 space dimensions.   We will
come to that model in section \ref{hmg}.

\section{Lecture Two}

\subsection{Chern-Simons Effective Action}

Today we will begin with an introduction to some aspects of the integer quantum Hall effect.   First I just want to explain from
the point of view of effective field theory why there is an integer quantum Hall effect in the first place.    We consider a material that not
only is an insulator, but more than that has no relevant degrees of freedom -- not even topological ones -- in the sense that its interaction
with an electromagnetic field can be described by an effective action for the $U(1)$ gauge field $A$ of electromagnetism only, without any additional degrees of
freedom.  (This would certainly not be true in a conductor, whose interaction with an electromagnetic field cannot be described without
including the charge carriers in the description, along with $A$.  But more subtly, as we will discuss, it is not true in a fractional quantum Hall system,
whose effective field theory requires topological degrees of freedom coupled to $A$.)

In a $3+1$-dimensional material with no relevant degrees of freedom, the effective action for the electromagnetic field can have all sorts of terms
associated to various familiar effects.   For example, ferromagnetism and ferroelectricity correspond to terms in the effective action
that are linear in $\vec E$ or $\vec B$
\be\label{dofof}I'=\int_{W_3\times \RR} \left(\vec a\cdot \vec E+\vec b\cdot \vec B\right).\ee
(Here $W_3$ is the spatial volume of the material and $\RR$ parametrizes the time, so the ``world-volume'' of the material is $M_4=W_3\times \RR$.)   Similarly, electric and magnetic susceptibilities
correspond to terms bilinear in $\vec E$ or $\vec B$:
\be\label{moof}I''=\int_{W_3\times \RR}\left(\alpha_{ij} E_iE_j+\beta_{ij}B_iB_j\right).\ee
And so on.

All these terms are manifestly gauge-invariant in the sense that they are integrals of gauge-invariant functions:  the integrands
are constructed only from $\vec E$ and $\vec B$ (and possibly their derivatives).   In $2+1$ dimensions, there is a unique term that is 
gauge-invariant but does {\it not} have this property.    This is the Chern-Simons coupling
\be\label{worf}\ICS=\frac{1}{4\pi}\int_{M_2\times \RR}\d^3x \epsilon^{ijk}A_i\partial_j A_k.\ee
The density $ \epsilon^{ijk}A_i\partial_j A_k$ that is being integrated is definitely not gauge-invariant, but the integral is gauge-invariant up to
a total derivative.   In fact, under
$A_i\to A_i+\partial_i\phi,$
we have
\be\label{omp} \epsilon^{ijk}A_i\partial_j A_k\to \epsilon^{ijk}A_i\partial_j A_k +\partial_i\left(\epsilon^{ijk}\phi\partial_jA_k\right).\ee

Roughly speaking, this shows that $\ICS$ is gauge-invariant, but we have to be more careful because of electric charge quantization.
If the quantum of electric charge is carried by a field $\psi$ of charge 1 transforming as
\be\label{perrf}\psi\to e^{\i\phi}\psi,\ee
then we should consider $\phi$ to be defined only modulo $2\pi$:
\be\label{eqr}\phi\cong\phi+2\pi.\ee
Given this fact, the previous proof of gauge-invariance of $\ICS$, in which $\phi$ was assumed to be single-valued, is not quite correct. 
We will be more careful in a moment.

Before I go on, though, I want to point out that logically, one could consider a theory in which one is only allowed to make a gauge
transformation $A_i\to A_i+\partial_i\phi$ with a single-valued $\phi$.    But that theory is not  the real world.    Dirac showed
that the Schrodinger equation of electrons, protons, and neutrons can be consistently coupled with magnetic monopoles, and that this consistency
is only possible because the Schrodinger equation is invariant under gauge transformations in which $e^{\i\phi}$ is single-valued although $\phi$
is not.    This is needed to make the Dirac string unobservable (fig. \ref{DString}).

  \begin{figure}
 \begin{center}
   \includegraphics[width=2.5in]{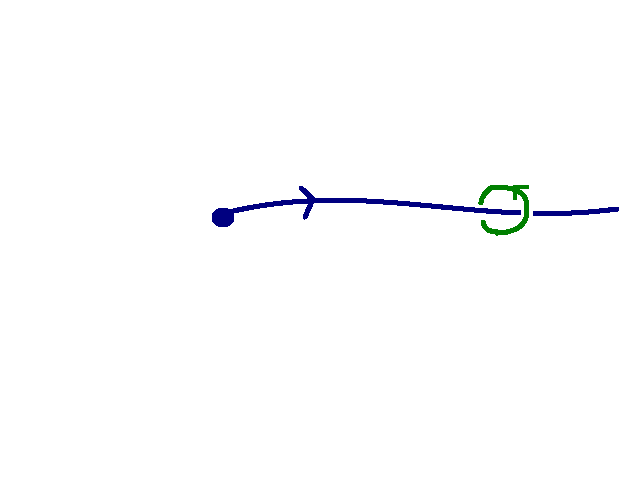}
 \end{center}
\caption{\small  The Dirac string emanating from a magnetic monopole is unobservable because the laws of nature are invariant
under gauge transformations in which $e^{\i\phi}$ is single-valued, but $\phi$ is not.}
 \label{DString}
\end{figure}
  
Anyway our microscopic knowledge that the Schrodinger equation is invariant under any gauge transformation such that $e^{\i\phi}$ is
single-valued (even if $\phi$ is not single-valued) implies constraints on the effective action that we would not have without that knowledge.
  We want to understand those constraints. 

\subsection{Quantization Of The Chern-Simons Coupling}

To do this, we will consider the following situation: we take our two-dimensional material to be a closed two-manifold, for instance $S^2$,
and we will take ``time'' to be a circle $S^1$ of circumference $\beta$.  (For example, we might be computing $\Tr\,e^{-\beta H}$.) 
Thus we consider a material whose ``worldvolume'' is $M_3=S^2\times S^1$ (fig. \ref{ST}).

\begin{figure}
 \begin{center}
   \includegraphics[width=2.5in]{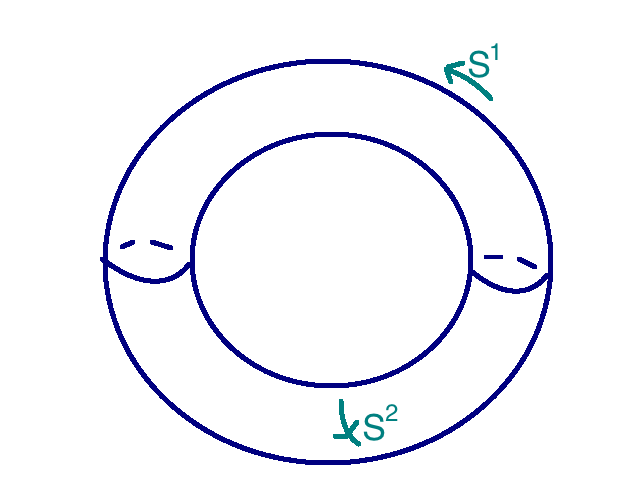}
 \end{center}
\caption{\small  A $2+1$-dimensional manifold $S^2\times S^1$ that is used in analyzing the gauge-invariance of the Chern-Simons action.}
 \label{ST}
\end{figure}

One might not be able to engineer this situation in the real world, but it is clear that the Schdrodinger equation makes sense in this situation. 
So we can consider it in deducing constraints on the effective action that can arise from the Schrodinger equation.

The gauge field that we want to consider on $M_3=S^2\times S^1$ is as follows.
We place a unit of Dirac magnetic flux on $S^2$:
\be\label{unitf}\int_{S^2}\d x_1 \d x_2\,\,\frac{F}{2\pi}=1.\ee
(This is the right quantum of flux if the covariant derivative of the electron is
$D_i\psi=(\partial_i-iA_i)\psi$, meaning that I am writing $A$ for what is often called $eA$.  This lets us avoid factors of $e$ in many formulas.)

And we take a constant gauge field in the $S^1$ or time direction:
\be\label{nif}A_0=\frac{s}{\beta},\ee
with constant $s$.      For this gauge field,
one can calculate\footnote{This is actually a slightly tricky calculation, because in the presence of nonvanishing magnetic flux on $S^2$, the
gauge field $A_i$ has a Dirac string singularity.  A safe way to do the calculation is to compute the derivative of $\ICS$ with respect to $s$,
using the fact that in any infinitesimal variation of $A$, one has $\delta \ICS=(1/4\pi)\int \epsilon_{ijk}\delta A_i F_{jk}$, which is written only
in terms of gauge-invariant quantitities $F_{jk}$ and $\delta A_i$.  Evaluating this formula for the case that $\delta A_i=\partial A_i/\partial_s=\delta_{i0}$ and that there is one unit of magnetc flux on $S^2$,
one finds that $\partial\ICS/\partial s=1$.  Using also the fact that $\ICS$ vanishes at $s=0$, one arrives at eqn. (\ref{nifty}).}
\be\label{nifty}\ICS= \frac{1}{4\pi}\int_{M_3=S^2\times S^1}\d^3x \epsilon^{ijk}A_i\partial_j A_k=s.\ee

 Note that the holonomy of $A$ around the ``time'' circle is
\be\label{wiff}\exp\left(i\int_0^\beta A_0\d t \right)=\exp\left(i\int_0^\beta (s/\beta)\d t\right)=\exp(is).\ee
The gauge transformation
\be\label{rif}\phi= \frac{2\pi t}{\beta},\ee 
which was chosen to make $e^{i\phi}$ periodic, acts by
\be\label{prif}s\to s+2\pi \ee
and so leaves the holonomy invariant.   
(This must be true, because with my normalization of $A$, this holonomy is the phase factor when an electron is parallel-transported
around the circle and so is physically meaningful.) 

So we have in this example $\ICS=s$, and a gauge transformation can act by $s\to s+2\pi$. 
Thus $\ICS $ is not quite gauge-invariant; it is only gauge-invariant mod $2\pi$.
   Here we must remember what is essentially the same fact that was exploited by Dirac
in his theory of the magnetic monopole.   The classical action $I$ enters quantum mechanics only via a factor $\exp(\i I)$
in the Feynman path integral (or $\exp(\i I/\hbar)$ if one restores $\hbar$), so it is enough if $I$ is well-defined and gauge-invariant mod $2\pi\Z$.
  Since $\ICS$ is actually gauge-invariant mod $2\pi \Z$ (we showed this in an example but it is actually true in general), it can
appear in the effective action with an integer coefficient:
$$I_\eff=k\ICS+\dots.$$

\subsection{Quantization Of The Hall Conductivity}

The point of this explanation has been to explain why $k$ has to be an integer -- sometimes called the ``level.''   The fact
that $k$ is an integer gives a macroscopic explanation of the quantization of the Hall current.    Indeed for any material whose interaction
with an electromagnetic potential $A$ is governed by an effective action $I_\eff$, the induced current in the material is
\be\label{curr}J_i=-\frac{\delta I_\eff}{\delta A_i}. \ee 
We are interested in the case that
\be\label{zurr}I_\eff=k\ICS=\frac{k}{4\pi}\int_{M_3}\d^3x \epsilon^{ijk}A_i\partial_jA_k.\ee

\begin{figure}
 \begin{center}
   \includegraphics[width=2.5in]{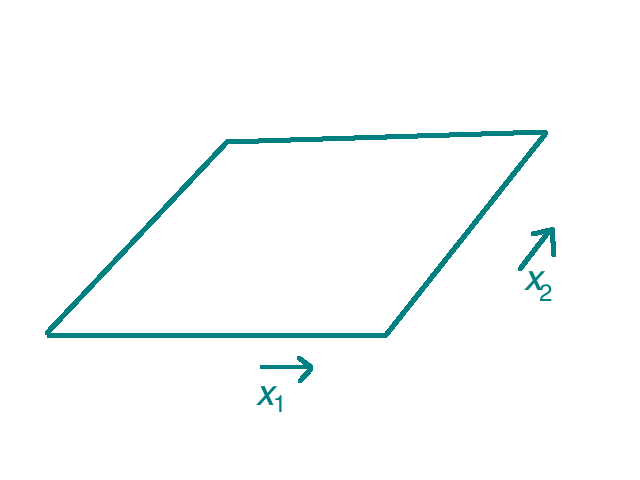}
 \end{center}
\caption{\small A two-dimensional sample.}
 \label{SeT}
\end{figure}
  Let us consider a material sitting at rest at $x_3=0$ and thus parametrized by $x_1,x_2$ (fig. \ref{SeT}).
  The current in the $x_2$ direction is
\be\label{murr}J_2=-\frac{\delta I_\eff}{\delta A_1}=\frac{kF_{01}}{2\pi}=\frac{kE_1}{2\pi}. \ee 
This is called a Hall current: an electric field in the $x_1$ direction has produced a current in the $x_2$ direction.   The Hall current
has a quantized coefficient $k/2\pi$ (usually called $ke^2/h$; recall that my $A$ is usually called $eA$ and that I set $\hbar=1$ so $h=2\pi$), where the quantization follows
from the fact that $\ICS $ is not quite gauge-invariant.

One may wonder ``How then can one have a fractional quantum Hall effect?''    I will give a short answer for now, postponing  more detail
for Lecture Three.\footnote{For a useful introduction to the very rich subject of effective field theories
 of the fractional quantum Hall effect, with references to the literature up to that time, see \cite{Wen}.}    One cannot get an integer quantum Hall effect in a description in which
$A$ is the only relevant degree of freedom.    However, from a macroscopic point of view, this can happen in a material that generates
an additional ``emergent'' $U(1)$ gauge field $a$ that only propagates in the material.   We normalize $a$ so that it has the same flux quantum\footnote{In
the context of condensed matter physics, it is unphysical to assume that an emergent gauge field $a$ gauges a noncompact gauge group $\RR$
rather than the compact gauge group $U(1)$.  In that case, there would be an exactly conserved current ${\mathbf j}_i=\frac{1}{2}\epsilon_{ijk}f^{jk}$ and
correspondingly, the space integral ${\mathbf q}=\int\d^2x \,{\mathbf{j}}^0=\int\d^2x\, f_{12}$ would be an exactly conserved quantity.  There
is no exactly conserved quantity in condensed matter physics that is a candidate for $\mathbf q$.  If the emergent gauge group is $U(1)$ rather than
$\RR$, then ``monopole operators'' can be added to the Hamiltonian, breaking the conservation of $\mathbf q$.  Because the emergent gauge group is
$U(1)$, there is a nontrivial Dirac flux quantum, and we normalize $a$ so that this quantum is $2\pi$.   Accordingly, the parameter $r$ in 
eqn. (\ref{urr}) must be an integer.}
$2\pi$
as $A$. 
We will write $f_{ij}=\partial_ia_j-\partial_ja_i$ for the field strength of $a$.
    An example   of a gauge-invariant effective action that
leads to a fractional quantum Hall effect is
\be\label{urr}I_\eff=\frac{1}{2\pi}\int_{M_3}\d^3x\,\,\epsilon^{ijk}A_i\partial_j a_k -\frac{r}{4\pi}\int_{M_3}\d^3x\,\,\epsilon^{ijk}a_i\partial_j a_k . \ee

An oversimplified explanation of why this gives a fractional quantum Hall effect is the following.   One argues that as $a$ appears
only quadratically in the effective action, one can integrate it out using its equation of motion.    This equation is
\be\label{yre}f=\frac{1}{r}F,\ee
implying that up to a gauge transformation $a=A/r$.  Substituting this in $I_\eff$, we get an effective action for $A$ only that describes
a fractional quantum Hall effect:
\be\label{zyre}I'_\eff=\frac{1}{r} \ICS(A)=\frac{1}{r}\frac{1}{4\pi}\int_{M_3}\d^3x \,\,\epsilon^{ijk}A_i\partial_jA_k.\ee

Here $1/r$ appears where $k$ usually does, and this suggests that the Hall conductivity in this model is $1/r$.  That is correct.   But there 
clearly is something wrong with the derivation
because the claimed answer for the effective action 
$I'_\eff=(1/r)\ICS(A)$ does not make sense as it violates gauge invariance. 
The mistake is that in general, as $F$ may have a flux quantum of $2\pi$, and $f$ has the same allowed flux quantum (otherwise the action we assumed
would not be gauge-invariant),  for a given $A$ it is not possible to solve the equation 
(\ref{yre})
for $a$.    Thus, it is not possible to eliminate $a$ from this system and give a description in terms of $A$ only.   The reason
that ``integrating out $a$'' gives the right answer for the Hall current is that this procedure is valid locally and this is enough to determine
the Hall current.  The system has more subtle properties (fractionally charged quasiparticles and topological degeneracies)
that can only be properly understood in the description with $a$ as well as $A$.  An introduction to those properties will be given in Lecture Three.

\subsection{Relation To Band Topology}

Going back to a theory that can be described in terms of $A$ only, we have then an integer $k$ in the macroscopic description. 
But there is also an integer in the microscopic description of a band insulator, the TKNN invariant \cite{TKNN}.   It
arises as follows.    We consider a crystal with $N$ bands, of which $n$ are filled.     We assume the system is completely gapped,
 for all values of the momentum.  As we learned yesterday, in a 2d system it is generic to have no band crossings.
 
 We are in the same situation as in our discussion yesterday of Weyl semimetals, except that there are no band-crossing points, so we work
 over the whole Brillouin zone $\B$, without removing anything.   As we are in two-dimensions, $\B$ is a two-torus. 
 At momentum $p$, let $\H_p\cong \C^N$ be the full space of all states, and $\H'_p$ the subspace of filled levels.    We can regard
 $\H_p$ as a rank (or dimension) $N$ ``trivial bundle'' over $\B$ and  $\H'_p$ as a ``subbundle'' of rank $n$.    The integer we want, which
 we will call $k'$, is
 the first Chern class $c_1(\H'_p)$, integrated over $\B$.   In terms of the Berry connection $\A$ on the filled bands
 that we discussed yesterday, whose curvature we call $\F$,
 this integer is 
 $$k'=c_1(\H'_p)=\int_\B \frac{\Tr\,\F}{2\pi}. $$  
 
 The basic claim of TKNN is that $k'$, the flux of the Berry connection, is the same as $k$, the coefficient of the quantum Hall
 current.    The original proof was based on literally just calculating the current from first principles in terms of a matrix element
 in the fermion ground state -- which is written as an integral of single particle matrix elements over the Brillouin zone.   I want to explain
 a different viewpoint that will emphasize that $k'$ is not just a band concept but can be defined in the full many-body theory.  (In Lecture Three,
 I will describe another approach essentially due to Haldane to the relation $k=k'$.)
  
 We consider a finite sample, say on an $n_1\times n_2$ lattice (fig. \ref{Rectangular2})
  for very large $n_1,n_2$,
where I will take lattice constants $a_1, a_2$ in the two directions.  Thus the physical size of the lattice is $L_1\times L_2$, with
$L_1=n_1a_1,$ $L_2=n_2a_2$.  We assume periodic boundary conditions, maintaining the lattice
translation symmetries.   However, for the finite system, the momenta take discrete values
\begin{align}\label{zudo}\notag p_1&=\frac{2\pi s_1}{L_1}, ~~0\leq s_1\leq n_1-1\cr
p_2 &=\frac{2\pi s_2}{L_2}, ~~0\leq s_2\leq n_2-1.\end{align}
  The ground state of the finite system is of course obtained by filling all of the states in the first  $n$ bands with these values of the momenta.

  \begin{figure}
 \begin{center}
   \includegraphics[width=2.5in]{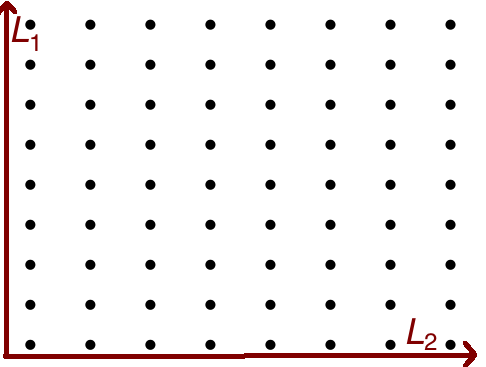}
 \end{center}
\caption{\small A periodic lattice of size $n_1\times n_2$, with lattice constants $a_1,a_2$ in the two directions.  The physical
size of the lattice is $L_1\times L_2$ with $L_1=n_1a_1, \,L_2=n_2a_2$.}
 \label{Rectangular2}
 \end{figure}
 
 Now, however, we turn on a background electromagnetic vector potential that is chosen such that the magnetic field vanishes, but an electron
going all the way around the $x_1$ direction or the $x_2$ direction picks up a phase: 
$$A_1=\frac{\alpha_1}{L_1}, ~~  A_2=\frac{\alpha_2}{L_2}. $$
  The phase picked up by an electron going around the $x_1$ (or $x_2$) direction is $\exp(i\alpha_1)$ (or $\exp(i\alpha_2)$)
and up to a gauge transformation the range of these parameters  is
$$0\leq \alpha_1 ,\,\,\alpha_2\leq 2\pi.$$      

From the point of view of band theory, the effect of turning on the parameters $\alpha_1,$ $\alpha_2$ is just to shift the momenta of
the electrons, which become
$$p_1=\frac{2\pi s_1+\alpha_1}{L_1}, ~~0\leq s_1\leq n_1-1$$
$$p_2=\frac{2\pi s_2+\alpha_2}{L_2}, ~~0\leq s_2\leq n_2-1.$$
  This actually shows that the spectrum is invariant under a $2\pi$ shift of $\alpha_1$ or of $\alpha_2$ (up to an integer shift of $s_1$ or $s_2$).
 For any $\alpha_1,\alpha_2$, from the point of view of band theory, the ground state is found by filling all states in the first $n$ bands 
with these shifted values of the momenta.

Now we think of the parameters $\alpha_1$, $\alpha_2$ as parameters that are going to vary adiabatically.   Since they are each
defined mod $2\pi$, they parametrize a torus that I will call $\widehat \B$.   ($\widehat \B$ can be viewed as a 
sort of rescaled version of the Brillouin
zone $\B$.)     Since Berry's construction is universal for adiabatic variation of parameters, 
we can construct a Berry connection $\widehat {\A}$ over 
$\widehat \B$,
with curvature $\widehat{ \F}$.    $\widehat {\A}$ is a connection that can be used to 
transport the ground state as the parameters $\alpha_1,\alpha_2$
are varied.    All we need to know to define it is that the ground state is always nondegenerate as $\alpha_1,\alpha_2$ are varied.  
We do not need to assume a single-particle picture (i.e. band theory).   But I should say that for the conclusions we draw to be useful,
at least in the form I will state, we need the gap from the ground state to be independent of $L_1,L_2$ as they become large. 
 (Otherwise in practice our measurements in the lab may not be adiabatic.   The stated assumption is not true for a fractional quantum Hall system, as we will
 discuss in Lecture Three.)
 
 Using the Berry connection over $\widehat B$, we can define an integer:
$$\widehat k' =\int_{\widehat B}\d\alpha_1\,\d\alpha_2\,\,\frac{\widehat \F}{2\pi}. $$
  But I claim that this is the same as the integer $k'$ defined in band theory:
$$k'=\widehat k'.$$

The reason that this is useful is that the definition of $\widehat k'$ is more general.    To define $k'$, we assume band theory -- that is,
a single-particle description based on free electrons.    The definition of $\widehat k'$ assumes much less.

  \begin{figure}
 \begin{center}
   \includegraphics[width=2in]{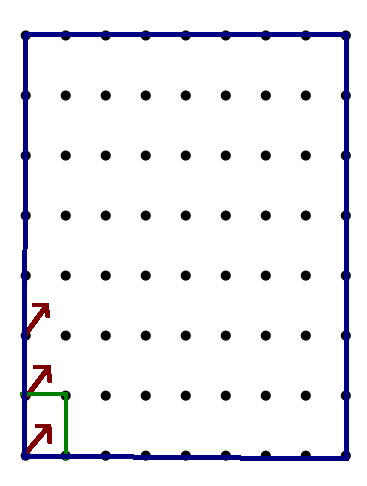}
 \end{center}
\caption{\small  The large rectangle (in which opposite sides should be identified) represents the momentum space Brillouin zone $\B$ that  parametrizes
the electron momentum.
The small rectangle in the lower left (with again opposite sides identified) represents the macroscopic Brillouin zone $\widehat\B$ that parametrizes the
shift angles $\alpha_1$, $\alpha_2$.  The black dots are the allowed values of the momentum for a finite lattice of size $n_1\times n_2$ (drawn here
for $n_1=8$, $n_2=7$).  Turning
on $\alpha_1$ and $\alpha_2$ shifts the allowed momenta as indicated by the arrows.  In particular, the point at the lower left of the picture
may be moved anywhere in the small rectangle that represents $\widehat \B$.
}
 \label{ShiftedLattice}
\end{figure}

\subsection{Proof Of The Equivalence}

To understand why $k'=\widehat k'$, I have drawn in fig. \ref{ShiftedLattice}
 the discrete points in the Brillouin zone that obey the finite volume condition.   
The parameters $\alpha_1,\alpha_2$ parametrize one of the little
rectangles in the picture, say the one at the lower left. 

To compute $k'$, we integrate over $\B$, the full Brillouin zone.   To compute $\hat k'$, we integrate over the little rectangle,
but for each point in the little rectangle, we sum over the corresponding shifted momenta.
These are two different ways to organize the same calculation, so $\hat k'= k'.$

So instead of proving the original TKNN formula $k=k'$, it is equivalent to prove that $k=\hat k'$.    This has the following advantage:
$\hat k'$ is defined in terms of the response of the system to a changing electromagnetic vector potential $A$, so we can determine 
$\hat k'$ just from a knowledge of the effective action for $A$.

As practice, before determining the Berry connection for $A$, I am going to determine the Berry connection for an arbitrary dynamical
system with dynamical variables $x^i(t)$.    You can think of $x^i(t)$, $i=1,\dots,3$  as representing the position coordinates
of a particle, but they really could be anything else (for example $x^i(t)$ could have $3N$ components representing the positions of $N$
particles).   Regardless, we assume an action
\be\label{merrof}I=\frac{1}{2}\int\d t\, g_{ij}(x)\frac{\d x^i}{\d t}\frac{\d x^j}{\d t}+ \int \d t \AA_i(x) \frac{\d x^i}{\d t}-\int\d t V(x)+\dots.\ee
(There might be higher order terms but it will be clear in a moment that they are not important.) 
We shall compute the Berry connection in the space of semiclassical states of zero energy, a condition that we satisfy by
imposing the condition $V(x)=0$.   (This semiclassical approximation is valid in our problem because we do not need to treat
the electromagnetic vector potential $A$ quantum mechanically.  We can view it as a given external field.)

 Setting $V=0$ means that we will evaluate the Berry phase not for all values of $x$ but only
for values of $x$ that ensure $V(x)=0$. So we  drop the $V(x)$ term from the action, and only carry out transport in the subspace of the configuration
space with $V=0$.   In adiabatic transport, we can also ignore the term 
\be\label{monkey}I_{\mathrm{kin}} =\frac{1}{2}\int\d t\, g_{ij}(x)\frac{\d x^i}{\d t}\frac{\d x^j}{\d t}\ee
in the action, and any other term with two or more time derivatives.    That is because if we transport from a starting point $p$
to an ending point $p'$ in time $T$, the derivative $\d x^i/\d t$ is of order $1/T$, and $I_{\mathrm{kin}}\sim 1/T$.  In the adiabatic limit,
$T\to \infty$ and this vanishes.

  \begin{figure}
 \begin{center}
   \includegraphics[width=2.5in]{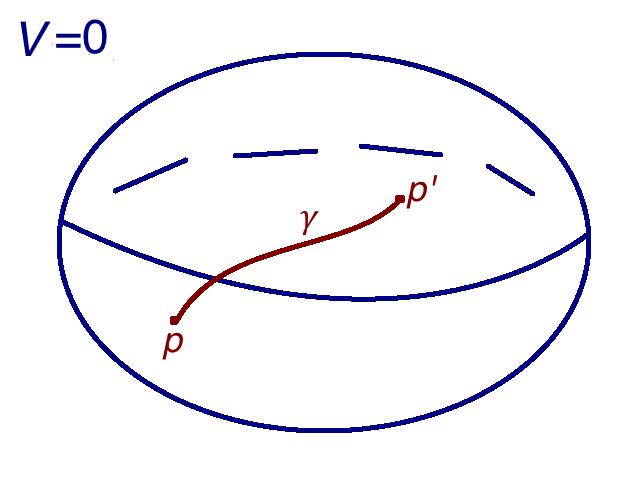}
 \end{center}
\caption{\small  Propagation of a quantum particle between two points $p$ and $p'$ in the configuration space.  We assume that the
propagation occurs within the subspace with $V=0$, represented here as a two-sphere.  The propagating particle acquires a phase $\exp(\i I/\hbar)$, where
$I$ is the classical action for the chosen trajectory.}
 \label{traj}
\end{figure}

So the only term in the action that we need to keep is the term with precisely one time derivative:
\be\label{wonkey}I'=\int\d t \,\AA_i(x)\frac{\d x^i}{\d t}=\int_p^{p'}\AA_i(x)\d x^i.\ee
  As I have indicated, this term depends only on the path followed from $p$ to $p'$, and not on how it is parametrized.
  Now remember that the phase that a quantum particle acquires in propagating from $p$ to $p'$ along a given trajectory
is $e^{\i I/\hbar}$, where $I$ is the action for that trajectory (fig. \ref{traj}).  For us, in units with $\hbar=1$,
 this phase is just $\exp\left(\i\int_\gamma\AA_i \d x^i\right)$.

But the connection which on parallel transport along a path $\gamma$ gives a phase
$\exp\left(\i\int_\gamma\AA_i \d x^i\right)$ is just $\AA$.    What we have learned, in other words, is that for a system
in which a quantum ground state can be considered to be equivalent to a classical ground state, the Berry connection is
just the classical connection $\AA$ that can be read off from the classical action.

For the electromagnetic field in our problem, the action is
\be\label{elfi}I=\frac{1}{2e^2}\int_{\RR^{3,1}}\d^3x \d t \left(\vec E^2 - \vec B^2\right) +\frac{k}{4\pi}\int_{W_3}\d^2x\d t \,\epsilon^{ijk}A_i\partial_j A_k+\dots  .\ee
  We assume, for example, periodic boundary conditions with very long periods $L_1,L_2$ in the two directions that are filled by our quantum Hall sample.  (It doesn't matter if we assume periodic boundary conditions in the third direction.)   A classical state of zero energy
is labeled by the two angles $\alpha_1$, $\alpha_2$ that were introduced earlier.
  To compute the Berry phase, we are supposed to substitute this formula in the action and keep only the part of the action
that has precisely 1 time derivative.   This comes only from the Chern-Simons term.  

After integration over $x_1$ and $x_2$,  the relevant part of the action is just
\be\label{mizz}I'=-\frac{k}{2\pi}\int\d t \,\alpha_1 \frac{\d \alpha_2}{\d t}.\ee
  From this, we read off the Berry connection
\be\label{roff} \nabla\equiv\left(\frac{D}{D\alpha_1},\frac{D}{D\alpha_2}\right)=\left(\frac{\partial}{\partial \alpha_1}, \frac{\partial}{\partial\alpha_2}+i\frac{k\alpha_1}{2\pi}\right),\ee
 and hence the Berry curvature
\be\label{coff}\hat\F_{\alpha_1\alpha_2}=-i\left[\frac{D}{D\alpha_1},\frac{D}{D\alpha_2}\right]=\frac{k}{2\pi}.  \ee
  (If we add to $I' $ a total derivative term $\int \d t\, \partial_t f(\alpha_1,\alpha_2)$, this will change the formula
for $\nabla$ but it will not change $\hat\F$.) 

We remember that the integer $\hat k'$ is supposed to be the integral of $\hat\F/2\pi$ over the Brillouin zone.  
We can now compute
\be\label{plizz}\hat k'=\int_0^{2\pi}\d\alpha_1\d\alpha_2\, \frac{\hat\F}{2\pi}= \int_0^{2\pi}\d\alpha_1\d\alpha_2\, \frac{k}{(2\pi)^2}=k.  \ee
  Thus we arrive at a version of the famous TKNN formula:  the coefficient $k$ of the quantum Hall current can be computed
as a flux integral of the Berry connection.  

\subsection{Edge States And Anomaly Inflow}

Yesterday, we explained why a purely 1d quantum electron gas cannot have an imbalance between left-moving and right-moving
electron excitations.   As a reminder, the reason was that in a periodic orbit, ``what goes up must come down''
(fig. \ref{Crossings}).   From a field theory point of view, this is needed because
right-moving gapless fermions without left-moving ones cannot be quantized in a gauge-invariant fashion.   There is a 1+1-dimensional
version of the 
Adler-Bell-Jackiw anomaly \cite{Adler,BJ}.

However, one of the hallmarks of a quantum Hall system is that on its boundary it has precisely such an imbalance.  
The reason that this must happen is that when we verified the invariance of the Chern-Simons action
\be\label{facs}k\ICS=\frac{k}{4\pi}\int_{M_2\times \RR}\d^3x \epsilon^{ijk}A_i\partial_j A_k\ee
 under a gauge transformation
$A_i\to A_i+\partial_i\phi$, we had to integrate by parts.    This integration by parts produces a surface term on the surface
of our material -- that is on $\partial M_2\times \RR$.  There is no way to cancel this failure of gauge invariance by adding
to the action a surface term supported on $\partial M_2\times \RR$.  
You can try to replace $\ICS$ by
\be\label{trytry}\ICS+\int_{\partial M_2\times \RR}\d t\,\d x\left(?????\right),\ee
where $?????$ is some polynomial in $A$ and its derivatives, but whatever you try will not work.  (I recommend this exercise.)   It is precisely
because the anomaly cannot be eliminated by adding some local interaction on the boundary that the anomaly is physically meaningful.

To cancel the anomaly, that is the failure of gauge invariance of $\ICS$ along the boundary, requires the existence on
the boundary of modes that are (1) gapless, so they cannot be integrated out to produce a local effective action for $A$ only, 
and (2) ``anomalous,'' that is they are not possible in a purely 1-dimensional system.  
What fills the bill is precisely what we found does not exist in a purely 1-dimensional system: ``chiral fermions,'' that is right-moving
gapless modes not accompanied by left-moving ones.  The fact that an anomaly in the boundary theory can be canceled by the existence
of a bulk interaction that is not gauge-invariant in the presence of a boundary is an illustration of the phenomenon of ``anomaly inflow'' \cite{CH}.

Since the failure of $k\ICS$ is proportional to $k$, the ``chiral asymmetry'' that is needed to cancel it is also proportional to $k$.
 In fact, the hallmark of an integer quantum Hall system with a Hall conductivity of $k$ is precisely that 
$$n_+-n_-=k$$
where $n_+$ and $n_-$ are the numbers of ``right-moving'' and ``left-moving'' gapless edge modes.   Instead of giving a technical
analysis of field theory anomalies to explain how this works, I will give a couple of possibly more physical explanations -- one today and one in Lecture
Three.  

\subsection{The Charge Pump}\label{tcp}

Today's explanation involves a version of the Thouless charge pump \cite{Thouless}.
Let us think of a quantum Hall system on the surface of a long cylinder.
  In fact for starters, think of an infinite cylinder (fig. \ref{Cylinder}).
    \begin{figure}
 \begin{center}
   \includegraphics[width=2.5in]{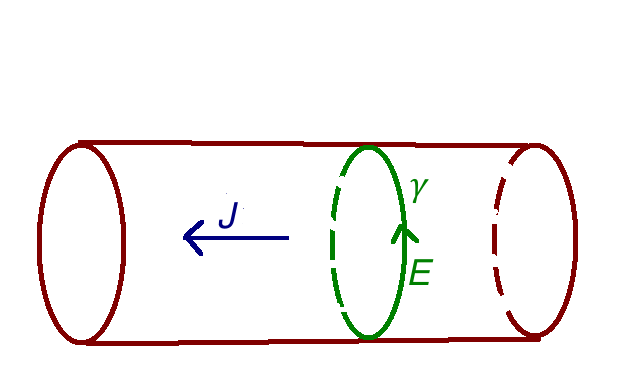}
 \end{center}
\caption{\small     Sketched here is a quantum Hall system supported on the surface of an infinite cylinder.  By varying $\oint_\gamma A\cdot \d\ell$, where $\gamma$
   is the indicated contour, one can induce a current that goes ``around'' the cylinder.  In a quantum Hall system, this will in turn induce an
   electric field ``along'' the cylinder, as indicated.}
 \label{Cylinder}
\end{figure}

We introduce the same sort of ``twist parameter'' $\alpha$ as before.    We can imagine that there is a magnetic flux
$\alpha$ through a solenoid inside the cylinder such that the magnetic field is 0 (or at least\footnote{In a conventional quantum Hall
system, there is a strong magnetic field in the sample -- that is in the cylinder -- but we can assume it to be independent of $\alpha$.
In a topological band insulator with $k\not=0$, the magnetic field can be assumed to vanish in the sample.  The latter case is
actually particularly natural for the discussion that follows.} independent of $\alpha$) in the cylinder itself
but $\oint_\gamma A\cdot \d\ell =\alpha.$
  Just as before, the parameter $\alpha$ is only gauge-invariant mod $2\pi$.

We adiabatically increase $\alpha$ from 0 to $2\pi$, with the scalar potential assumed to be 0.   Since the electric field
is then \be\label{zimmo}\vec E=\frac{\partial \vec A}{\partial t},\ee
increasing $\alpha$ turns on an electric field that goes ``around'' the cylinder.   But in the case of a quantum Hall system,
this drives a current that is perpendicular to $\vec E$, in other words the current flows ``along'' the cylinder. 
 The electrons therefore are pushed to the left (or right, depending on the sign of $k$).

An early explanation by Laughlin of the integer quantum Hall effect was the following.    We assume that when $\alpha=2\pi$,
the system returns to the same state that it was in at $\alpha=0$.  (This assumption is not valid for fractional quantum Hall systems, as explained
in section \ref{moref}.)   However, in the process, each electron may move $k$ steps to the left, for some integer $k$.   Notice
that since the cylinder has a finite circumference $S$, the number of electrons per unit length is finite and thus it makes sense  to say
that each one moves $k$ steps to the left, for some $k$.   This was interpreted as the basic integrality of the integer quantum 
Hall effect.   It does lead to the value $k/2\pi$ for the Hall conductivity.

  \begin{figure}
 \begin{center}
   \includegraphics[width=2.5in]{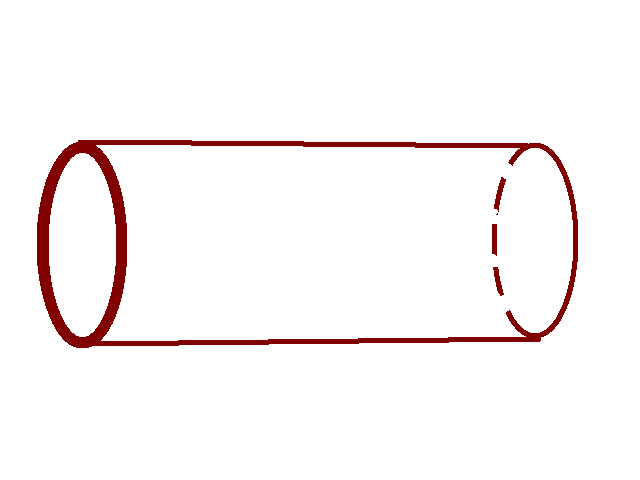}
 \end{center}
\caption{\small A semi-infinite cylinder with a boundary at the left end.}
 \label{Cylindrical4}
\end{figure}
Now let us consider a cylinder that is only {\it semi}-infinite, with a boundary at let us say the left end
(fig. \ref{Cylindrical4}).
 The same parameter $\alpha$ as before makes sense, and we can still adiabatically increase it by $2\pi$.    Since
 a quantum Hall system is gapped, if we make a measurement  far from the boundary, we will still see the same flux of valence electrons to the left as before,
 assuming that only valence bands (states below the fermi energy) are filled.
 
 But what happens to the electrons when they arrive at the left boundary?   A partial answer is that there are edge
 states, and electrons go from the valence bands to the edge states.   But this is not enough: since the boundary has finite
 length, only finitely many electrons can go into edge states (of reasonable energy).   What happens, at least in a topological
 band insulator (with finitely  many bands) in which there is an upper bound on the possible energy of an electron, is that as electrons
 flow in to the left from the valence bands (the bands below the usual $\veps_F$ in the bulk) they must eventually flow back out to the right
 in the conduction bands (the bands above the usual $\veps_F$).   Moreover, all this is happening continuously in energy
 so it must be  possible for an electron to evolve {\it continuously} from the valence bands in the bulk, to the conduction bands in the bulk,
 somehow passing through edge states.
 
   \begin{figure}
 \begin{center}
   \includegraphics[width=3in]{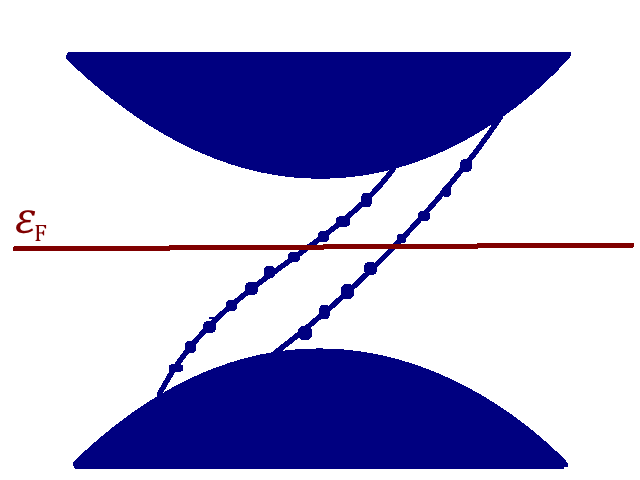}
 \end{center}
\caption{\small  Drawn here is the $\veps-p$ plane, where $\veps$ is the electron energy and $p$ is the electron momentum along the boundary
of a quantum Hall region.  The solid regions are projections of the valence bands and conduction bands to the $\veps-p$ plane.  (In these projections,
the component of electron momentum normal to the boundary is forgotten, as it is not a conserved quantity near the boundary.) Each curve
in the figure represents a right-moving edge mode that connects the projection of the bulk valence bands to the projection of the bulk conduction
bands.  (In the case shown, there are two such curves, corresponding to $k=2$.)
Such a  curve represents the energy-momentum relation of a family of edge-localized states; as in the discussion of Fermi arcs in Lecture One,
such a curve terminates at a point where edge-localization breaks down and the edge-localized state becomes indistinguishable form a bulk state.
The little beads represent, for some value of the parameter $\alpha$, allowed values of the boundary momentum of an edge state when the circumference $S$ of the boundary is finite.}
 \label{HalfBands}
\end{figure}
 
 \subsection{Joining Valence And Conduction Bands}
 
  The spectrum must therefore look something like  what is shown in fig. \ref{HalfBands}: there must be edge-localized states that
  can continuously leave the valence band, flow up through the Fermi energy, and eventually join the conduction bands.  In the limit that the circumference
  $S$ of the cylinder becomes infinite, the edge-localized states have a continuous spectrum, but for finite $S$, they have a discrete spectrum,
  as shown in the figure.  In the figure, each curve in the $\veps-p$ plane that connects the valence bands to the conduction bands
  represents a right-moving edge mode.  The little beads on the curve represent the allowed points on the curve for some large value of $S$.
    As we adiabatically increase $\alpha$, each little bead moves up along
 the curve and under $\alpha\to \alpha+2\pi$, each bead is shifted in position to the next one.    So under $\alpha\to \alpha+2\pi$,
 there is a net charge flow
 of 1 from the valence bands to the conduction bands for each right-moving edge mode.
 
     Recall that as we discussed in Lecture One,
 a 1d mode is rightmoving if $\d\veps/\d p>0$ at  $\veps=\veps_F$.    A left-moving mode has $\d\veps/\d p<0$ at $\veps=\veps_F$,
 and under $\alpha\to \alpha+2\pi$ produces a net charge flow of $-1$ from the valence band to the conduction band. 
 Thus with $n_+$ and $n_-$ as the numbers of right- and left-moving modes, the net charge flow under $\alpha\to\alpha+2\pi$
 is
 \be\label{trof}k=n_+-n_-.\ee
 
It remains to tie up some loose ends in this explanation.
 The  1d edge modes cannot be defined on the whole 1d Brillouin zone of the boundary (which is a circle) because then we would be stuck
 with the fact that in a periodic orbit ``what goes up must come down,'' leading to $n_+=n_-$.   The asymmetry comes from branches of
 edge mode that
 exist in only a finite
 range of momenta $p_-\leq p\leq p_+$.    What happens at the endpoints?    The answer is the same as it was in the somewhat
 similar example of Fermi arcs that we discussed yesterday.   The way that a family of edge-localized states can cease to exist at some momentum is by ceasing to be normalizable.    This happens when the edge state becomes indistinguishable from a bulk state.
 
  That is part of what makes it possible to have adiabatic transport from the valence bands  (the states normally filled)
  to the conduction bands (the states normally empty), through
  the edge states.   At the endpoint of the edge state spectrum, an edge state is indistinguishable from a bulk state.
 
 For all this to make sense, adiabatic transport must remove as many electrons from the boundary in the conduction bands
 as approach the boundary in the valence bands.  In other words, 
 the total Hall conductivity of the empty (conduction) bands  must be minus the Hall conductivity of
 the filled (valence) bands.  That is actually a property of the Berry connection.   Let $\A$ be the usual Berry connection
 for the filled bands and $\F$ the corresponding curvature; and similarly let $\A'$ and $\F'$ be the Berry connection and curvature of the
 empty bands.    Then $\Tr\,\F+\Tr\,\F'=0$, basically because for all bands together there is no Berry curvature.  (The sum $(\Tr\,\F+\Tr\,\F')/2\pi$
 would represent the first Chern class of all bands together, and this vanishes because all the electron states together form a trivial
 vector bundle over the Brillouin zone.)
 
 Indeed, the Hall conductivities of filled and empty bands are respectively
 \be\label{twobbb}\int_B\frac{\Tr \,\F}{2\pi},~~~~\int_B\frac{\Tr\,\F'}{2\pi}.\ee
   So the relation $\Tr\,\F+\Tr\,\F'=0$ means that conduction and valence bands have opposite Hall conductivities, and in our
 thought experiment, the flow of ``filled'' states (i.e., states that would be filled in the ground state on an infinite cylinder) 
 to the left equals the flow of ``empty'' states to the right.
 
 \subsection{More On The Fractional Quantum Hall Effect}\label{moref}
 
I would like to next explain the assertion that a fractional quantum Hall system does not return to its previous state under
 $\alpha\to \alpha+2\pi$.    We will use the same macroscopic model of a fractional quantum Hall system that we used before
 in terms of the electromagnetic vector potential $A$ and an emergent $U(1)$ gauge field $a$ that only exists inside the material:
 $$I_\eff=\frac{1}{2\pi}\int_{M_3}\d^3x\,\,\epsilon^{ijk}A_i\partial_j a_k -\frac{r}{4\pi}\int_{M_3}\d^3x\,\,\epsilon^{ijk}a_i\partial_j a_k . $$ 
 As in fig. \ref{Cylinder}, we consider a   cylindrical sample and define $\alpha=\oint_\gamma A$.
   First let us discuss how to characterize the state of the system for a given $\alpha$.
 
 In principle, $\alpha=\oint_\gamma A$ can be controlled by varying the magnetic flux threaded by the 
 cylinder.    But there is an analogous parameter $\hat\alpha=\oint_\gamma a$
 that cannot be controlled in that way. 
 Just like $\alpha$, $\hat\alpha$ is gauge-invariant mod $2\pi$.  
 
  What can we say about $\hat\alpha$?   Recalling that $F=\d A$, $f=\d a$ are the ordinary
 electromagnetic field strength and its analog for $a$, the classical field equation for this system is
 \be\label{classeq}rf=F.\ee
   In the limit of an infinite cylinder, $a$ can be treated classically.  (We postpone the more interesting case of a finite cylinder
 until Lecture Three.)
   In the gauge $A_0=a_0=0$, the equation $rf_{0i}=F_{0i}$ becomes
 $$r\frac{\d a_i }{\d t}=\frac{\d A_i}{\d t},$$
 and therefore
 $$r\frac{\d\hat\alpha}{\d t}=\frac{\d\alpha}{\d t}.$$
 
 Hence when we adiabatically increase $\alpha$ by $2\pi$, $\hat\alpha$ increases adiabatically by $2\pi/r$.    Since
 $\hat\alpha$ is gauge-invariant mod $2\pi$, the shift $\hat\alpha\to\hat\alpha+2\pi/r$ does not return the system to its original state.
   We need to take $\alpha\to \alpha+2\pi r$, and therefore the Hall conductivity can be smaller than its usual ``quantum''
 by a factor of $r$.
 
 Fig. 
\ref{HalfBands}
 still has some sort of analog,  but the edge states cannot be free electron states:    They have to be capable of
 transporting a fractional charge under $\alpha\to\alpha+2\pi$, and returning to their original state only under $\alpha\to\alpha+2\pi r$.

\subsection{More On Fermi Arcs}

Finally, we will take another look at the Fermi arcs that we discussed in Lecture One.  In Lecture One, we considered a model Hamiltonian
that is valid near a generic band-crossing point and did an explicit computation to show the appearance of edge-localized states.  
However, the original paper \cite{WTVS} predicting these states did not proceed by solving a model Schrodinger equation.   Rather
the result was deduced as follows from some of the things that we have explained today.   
 
First we recall the basic setup.   Weyl points arise at special points in the Brillouin zone at which valence and conduction
bands meet (fig. \ref{twoob}). 
 Near a boundary of a finite sample, only two of the three components of momentum are conserved.  So it is useful (fig. \ref{projection})
 to project the Brillouin zone and the bad points in it to two dimensions, ``forgetting'' the component of momentum that is not conserved.  It is important to remember that in a crystal, the momentum
  components, including the component that is being ``forgotten'', are periodic, and in particular the horizontal direction in the picture
  represents a circle $U\cong S^1$, though it is hard to draw this.
  
    \begin{figure}
 \begin{center}
   \includegraphics[width=2.5in]{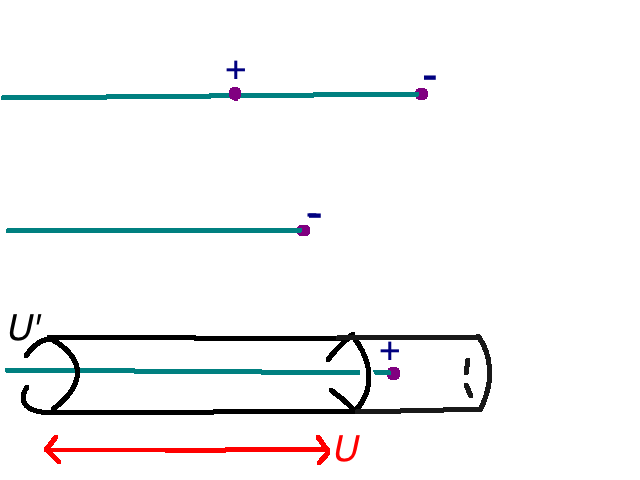}
 \end{center}
\caption{\small Projection of the bulk Brillouin zone to the boundary Brillouin zone that is appropriate for edge states. 
In the projection, the component of electron momentum normal to the boundary (here plotted horizontally) is forgotten.  This component
is parametrized by a circle $U$.  In the figure, we have drawn another circle $U'$ around the projection of one of the band-crossing points in the bulk.
The product $U\times U'$ is a two-torus.  This two-torus is used as the Brillouin zone of an auxiliary quantum Hall system that can be used
in analyzing the Fermi arcs.}
 \label{Special3}
\end{figure}
  Now  draw a little circle $U'$ around the projection of one of the bad points, as in fig. \ref{Special3}.    
  The product $U\times U'$ is a two-torus.
     We define an integer $k^*$ as the Berry flux through $U\times U'$:
     $$k^*=\int_{U\times U'}\d^2 p \,\frac{\Tr\,\F}{2\pi}. $$
   It receives a contribution of $1$ or $-1$ for each positive or negative  Weyl point enclosed by $U\times U'$. 
   So in the example drawn, $k^*=1$, but we would get $k^*=0$ or $k^*=-1$ if we take $U'$ to encircle one of the other two
   special points in the projection of fig. \ref{Special3}. 
   
   We have arranged so that the two-torus $U\times U'$ does not intersect any of the Weyl points.    So the restriction to $U\times U'$ of  the original
   3d band Hamiltonian on the 3d Brillouin zone $B$ is a  gapped  Hamiltonian $H^*$ parametrized by a two-torus $U\times U'$.    We can
   intepret $H^*$ as the band Hamiltonian of some 2d lattice system that has a Hall conductivity of $k^*$.    So as we have learned,
   $H^*$ has edge modes, equal in number to $k^*$,  that ``bridge the gap'' in energy between the filled and empty bands.
   This is sketched in fig. \ref{Special4}.
   
       \begin{figure}
 \begin{center}
   \includegraphics[width=2.5in]{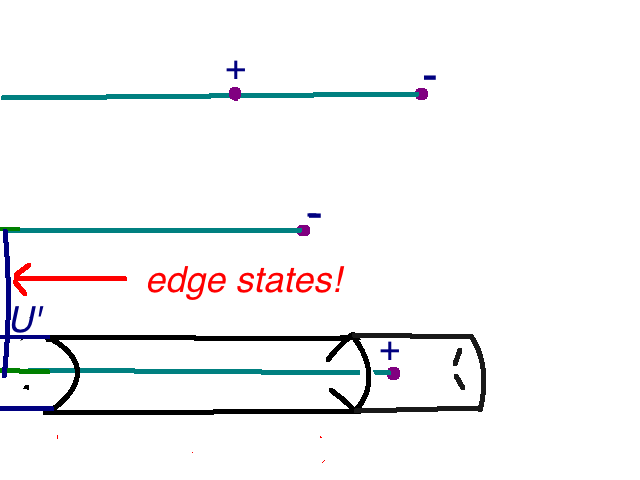}
 \end{center}
\caption{\small The band topology on $U\times U'$ is such that there must be edge states intersecting the circle $U'$ and connecting
the points in the boundary Brillouin zone that are projections of bulk band-crossing points with a nonzero net chirality.}
 \label{Special4}
\end{figure}
     
  So there have to be edge states that intersect $U'$ (the edge states are not labeled by $U$ since $U$ parametrizes the
  component of momentum that is not relevant to edge states).    Since we had a lot of freedom in the choice of $U'$, the spectrum
  of edge-localized states has to consist of arcs that join the projections of band-crossing points.     

       The auxiliary 2d quantum Hall system that was used in this argument does not have any simple relation, as far as I know,
       to the 3d Weyl semi-metal that we are studying.

\section{Lecture Three}

Today's lecture will concern three topics:  (1) more on the fractional quantum Hall effect; (2) another explanation of the edge modes in the integer
quantum Hall effect; (3) Haldane's model \cite{Haldane} of quantum Hall physics without an applied magnetic field.

\subsection{More On The Fractional Quantum Hall Effect}

As yesterday, we describe a fractional quantum Hall system macroscopically by
an effective action for the electromagnetic vector potential $A$ and an ``emergent'' $U(1)$ vector potential $a$ that only exists inside the fractional Hall material.
  We can take the effective action to be the sum of the bulk Maxwell action
\be\label{delf}\frac{1}{2e^2}\int_{\RR^{3,1}}\d^3x\d t\,\left(\vec E^2-\vec B^2\right)\ee
plus a term that ``lives'' in the material:
\be\label{welf}I_\eff=\int_{M_2\times \RR}\d^2x\d t\,\left(\frac{1}{2\pi}\epsilon^{ijk}A_i\partial_j a_k -\frac{r}{4\pi}\epsilon^{ijk}a_i\partial_j a_k\right) . \ee

For a first orientation, let us consider the interpretation of a ``quasiparticle'' that has a charge $q$ under $a$.  Here  $q$ must be an integer
since $a$ is a $U(1)$ gauge field.  If such a quasiparticle is present at rest at a point $x=x_0$ in $M_2$, 
then the part of the action that depends on $a$ acquires an extra term and becomes
\be\label{melf}\int_{M_2\times \RR}\d^2x\d t\,\left(\frac{1}{2\pi}\epsilon^{ijk}A_i\partial_j a_k -\frac{r}{4\pi}\epsilon^{ijk}a_i\partial_j a_k\right) +q\int\d t
\,a_0(x_0,t). \ee
The field equation for $a_0$ becomes
\be\label{pelf}\frac{F_{12}(x)}{2\pi}-\frac{rf_{12}(x)}{2\pi} +q\delta(x-x_0)=0.\ee

To solve this equation, we obviously need a delta function in $F_{12}$ and/or $f_{12}$.   But which?    In condensed
matter, a delta function in $f$ or $F$ is really an idealization of a very tiny flux tube.    Because $a$ and $f$ live only in two space dimensions,
a delta function in $f$ makes sense.  It represents a little flux tube supported near the point $x_0$ in the two-dimensional
 surface $M_2$ (fig. \ref{Deltao}).
The coefficient of the delta function determines the integral of $f$ over $M_2$.
  \begin{figure}
 \begin{center}
   \includegraphics[width=2.5in]{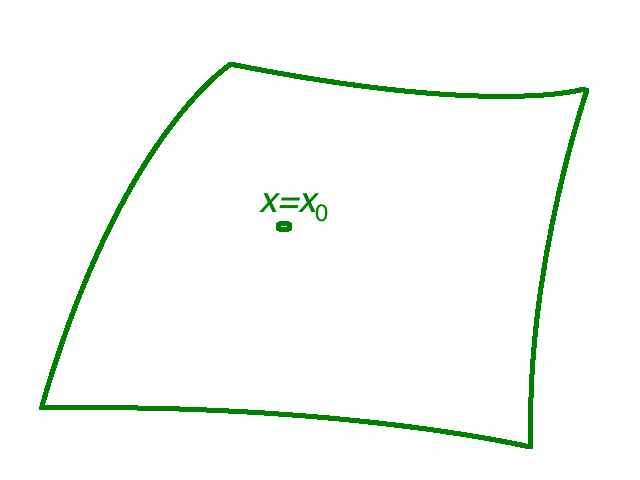}
 \end{center}
\caption{\small A flux tube of $f$ in a two-dimensional surface.  $f$ vanishes except very near the point $x=x_0$, and
has a nonzero integral.}
 \label{Deltao}
\end{figure}

Because $A$ lives throughout all of $3+1$ dimensional spacetime, such a delta function does not make sense for $F=\d A$.
 Of course, we can imagine a thin solenoid generating a flux tube of $F$, but this would extend into the third spatial dimension.
It would not represent a quasiparticle that lives in $M_2$.  Alternatively, we could consider a small electric current loop in $M_2$,
say of small radius $\rho$.
Such a current loop will create a magnetic dipole field in $3+1$ dimensions.  The magnetic field lines of a dipole form closed
loops and there is no net flux through $M_2$: 
if $F$ is the dipolar magnetic field of a current loop 
then $\int_{M_2}F=0$.  (This is true whether or not the current loop is in $M_2$.)
So after coarse-graining, a small current loop will give 0, not a delta function in $F$.
 
 So we have to solve the equation
 \be\label{lof}\frac{F_{12}(x)}{2\pi}-\frac{rf_{12}(x)}{2\pi} +q\delta(x-x_0)=0\ee
 with a delta function in $f_{12}$ and not in $F_{12}$, and hence near $x=x_0$,
 \be\label{qof}\frac{f_{12}(x)}{2\pi}=\frac{q}{r}\delta(x-x_0).\ee
 Now if we go back to the action
 \be\label{meff}I_\eff=\frac{1}{2\pi}\int_{M_2\times \RR}\d^2x\d t\,\,\epsilon^{ijk}A_i\partial_j a_k -\frac{r}{4\pi}\int_{M_2\times\RR}\d^3x\,\,\epsilon^{ijk}a_i\partial_j a_k ,\ee
 we see that the charge density $J_0=\delta I_\eff/\delta A_0$ of $A$ is
 \be\label{pefr}J_0=\frac{f_{12}}{2\pi}=\frac{q}{r} \delta(x-x_0).\ee
  Thus a quasiparticle with charge $q$ for $a$ has ordinary electric charge $q/r$ (in units of the charge of the electron).

 It is actually not necessary to go into so much detail to see that a  fractional quantum Hall system must have fractionally charged
 quasiparticles.    Return for a moment to an integer quantum Hall system:
 \be\label{tfor}I_\eff=\frac{k}{4\pi}\int \d^2x \d t \,\epsilon^{ijk}A_i\partial_j A_k,~~~k\in \Z.\ee
  The corresponding electric charge density is
\be\label{dfor}J_0=\frac{\delta I_\eff}{\delta A_0}=\frac{kF_{12}}{2\pi}.\ee 

  \begin{figure}
 \begin{center}
   \includegraphics[width=2.5in]{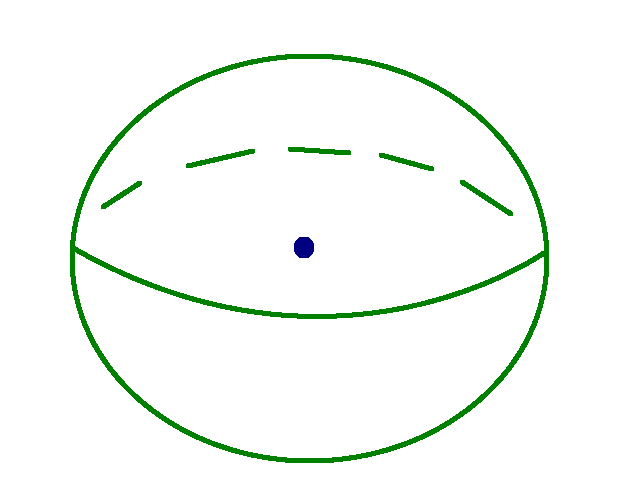}
 \end{center}
\caption{\small A spherical quantum Hall sample with a magnetic monopole inside.  The Chern-Simons effective action
predicts that if no quasiparticles are present, then the induced electric charge in the quantum Hall sample will equal $k$,
the coefficient of the Hall conductivity. The system can be everywhere near its ground state
 if $k$ is an integer, but if $k$ is a fraction, then fractionally charged quasiparticles will inevitably be present.}\label{Monopole}
\end{figure}

If we place a magnetic monopole with one Dirac quantum
\be\label{groof}\int \frac{F_{12}}{2\pi}=1\ee
inside a spherical sample (fig. \ref{Monopole}), then the Chern-Simons effective action predicts that
   this induces in the material a charge
 \be\label{droof}Q=\int_{M_2}  J_0=\int_{M_2} \frac{kF_{12}}{2\pi}=k,\ee
 and all is well if $k$ is an integer.
 But if the effective value of $k$ is not an integer, as in the case of the fractional quantum Hall effect, then there must be
 additional contributions to the electric charge in the form of fractionally charged quasiparticles that will appear somewhere
 on the surface of the material.  Indeed, consider a thought experiment in which we start with an isolated magnetic monopole
 in vacuum.  Then we bring in from infinity a large but finite collection of atoms and assemble them into a fractional quantum 
 Hall system in the form of a large sphere  surrounding the magnetic monopole, as in the figure.  We can do this making sure
 that the atoms are always far away from the monopole.  As electric charge is conserved and the vacuum is an insulator,
 the charge of the monopole does not change in this process.  The total electric charge of the spherical quantum Hall sample
 that surrounds the monopole will be an integer, since this system is ultimately made from a finite number of electrons, protons,
 and neutrons, each of integer charge.  The total electric charge of the system, however, will be the sum of the ``bulk'' contribution
 $k$ of eqn. (\ref{droof}) and a further contribution from any quasiparticles that may be present.  If $k$ is not an integer,
 then the spherical quantum Hall system that surrounds a magnetic monopole will have to contain fractionally charged quasiparticles.

 A full understanding of the fractional quantum Hall system requires treating $a$ quantum mechanically.      I will not attempt a complete explanation  in this lecture, but will explain a few basic points.
 
 First of all, part of the reason that we have gotten as far as we have without treating $a$ quantum mechanically is that so far
 we considered a fractional quantum Hall system on an infinite or semi-infinite cylinder such as that of fig. \ref{Cylindrical4},
  in which case a full quantum treatment is not necessary for many questions.
   For example, in section \ref{moref}, we treated $\hat\alpha=\oint_\gamma a$ as an arbitrary constant, rather than a quantum variable.   This
is possible on an infinite or semi-infinite cylinder, but in the case of a compact sample we do need to treat $a$ quantum mechanically. 
 
  In discussing
 the quantum mechanics of $a$, we will ignore $A$ and just study the purely $2+1$-dimensional problem:
 \be\label{loof}I_\eff=-\frac{r}{4\pi}\int_{M_2\times\RR}\d^3x\,\,\epsilon^{ijk}a_i\partial_j a_k . \ee     A noteworthy fact is that {\it there is no metric tensor in 
 sight}, and therefore what we are trying to describe is a ``topological quantum field theory.''    It will describe not particle excitations,
 but only the ``dynamics of the ground state(s)'' and topological properties of quasiparticles.   At long distances, many or most gapped quantum systems simply become trivial,  and usually we take this for granted as the long distance behavior of a gapped system.
 But more generally a gapped quantum system can lead at long distances to a nontrivial topological quantum field theory, and that is what
 happens in the case of the fractional quantum Hall effect.
 
   \begin{figure}
 \begin{center}
   \includegraphics[width=2.5in]{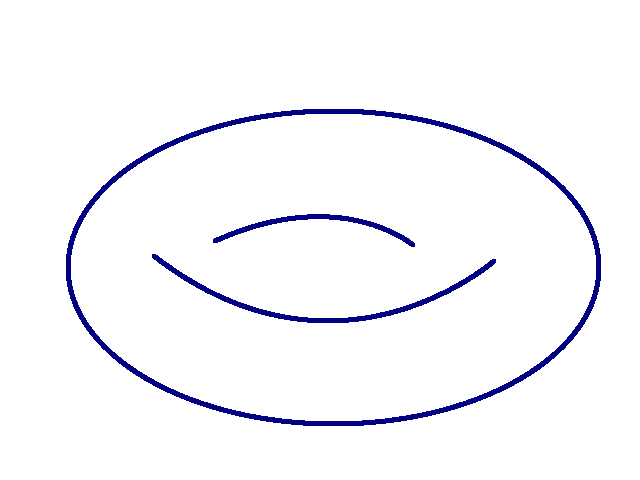}
 \end{center}
\caption{\small A two-dimensional torus.}
 \label{Torus}
\end{figure}
 A compact sample of nontrivial topology, such as the torus of fig. \ref{Torus}, is particularly interesting.
 We want to find the quantum states of the field $a$ quantized on such a manifold.

 The quantum states are supposed to make up a Hilbert space $\HH$.    $\HH$ is supposed to provide a representation of an algebra
 of quantum operators that is obtained, in some sense, by quantizing the space of classical observables.    In a gauge theory, we consider
 only the gauge-invariant classical observables.  So we should ask,
 ``What are the gauge-invariant classical observables that we can make from $a$?''   As soon as we ask this question, we run into
 the following fact.   A gauge-invariant local operator would have to be a polynomial in $f=\d a$ and its derivatives.   But the
 classical field equation of $a$ is
 \be\label{doof}f=0,\ee
 and therefore there are no local, gauge-invariant classical observables.
 
 However, there {\it are} gauge-invariant ``Wilson loop'' operators.  We pick a closed curve $\ell\subset M$ and define the ``Wilson loop
 operator''
 \be\label{ooff}W_s(\ell)=\exp\left(is\oint_\ell a\right),~~~s\in \Z.\ee  This operator is invariant under continuous deformations of $\ell$.
Here are two related explanations of this fact:  (a) This is true because $f=0$ so $W_s(\ell)$ can only see global information like Aharonov-Bohm phases; 
 (b) More generally, in any topological quantum field theory, diffeomorphisms are symmetries
 and any loop $\ell$ is equivalent by diffeomorphism to any nearby loop  to which $\ell$ can be deformed.

 The physical meaning of the Wilson loop operator $W_s(\ell)$ is that the amplitude for a process in which a quasiparticle of charge $s$
 propagates around a loop $\ell$ is proportional to a factor of $W_s(\ell)$.    If the loop $\ell$ can be continuously shrunk to a point
 without any singularity, then the operator $W_s(\ell)$ is trivial since the quasiparticle is not going anywhere.  ``Trivial'' means
 that in this case $W_s(\ell)$ is equal to 1 as an operator.    We are only interested in the case
 that this is not so.
 
  \begin{figure}
 \begin{center}
   \includegraphics[width=2.5in]{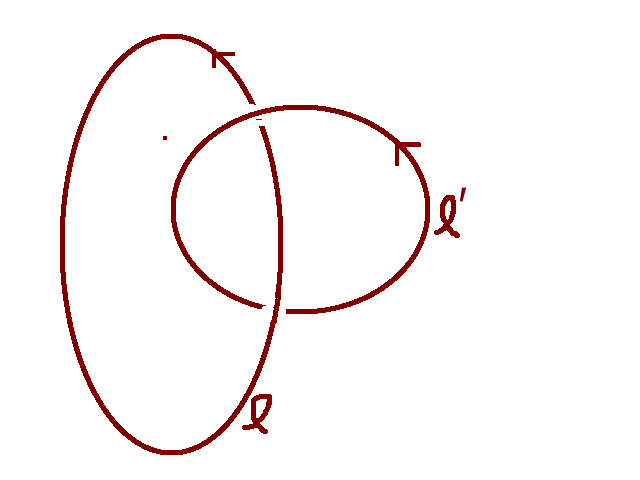}
 \end{center}
\caption{\small   Linking of two loops in $\RR^3$.}
 \label{Loops}
\end{figure}
 There are two possible sources of Aharonov-Bohm-like phases that a Wilson loop operator $W_s(\ell)$ might see. 
There may be  another similar operator $W_{s'}(\ell')$, where $\ell$ and $\ell'$ are ``linked'' and cannot be disentangled (fig. \ref{Loops}).
    (There will be a singularity if we try to pass $\ell$ through $\ell'$.)  
  This effect is associated to fractional statistics of the quasiparticles:  It means that the presence of one quasiparticle propagating around $\ell'$
  modifies the amplitude for a second quasiparticle to propagate around $\ell$ even if they are very far apart.
 Alternatively, and more like the classical Aharonov-Bohm idea, the loop $\ell$ might be ``noncontractible''  for topological reasons unrelated to the existence of other quasiparticles (fig. \ref{Torus2}).
 
   \begin{figure}
 \begin{center}
   \includegraphics[width=2.5in]{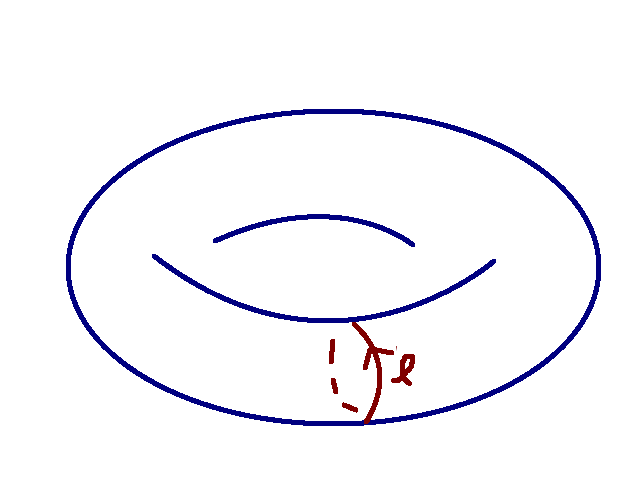}
 \end{center}
\caption{\small A noncontractible loop on a torus.}
 \label{Torus2}
\end{figure}

 If two such loops $\ell$ and $\ell'$ on the torus have a nonzero intersection number  $\ell\cap\ell'$ (fig. \ref{Torus3})  then -- as one can learn with the help of the classical Poisson brackets or quantum canonical commutators -- the corresponding
   Wilson operators do not commute.    They obey
   \be\label{grooff}W_s(\ell) W_{s'}(\ell') =\exp(2\pi i ss'\ell\cap\ell'/r) W_{s'}(\ell')W_s(\ell).\ee
   
    \begin{figure}
 \begin{center}
   \includegraphics[width=2.5in]{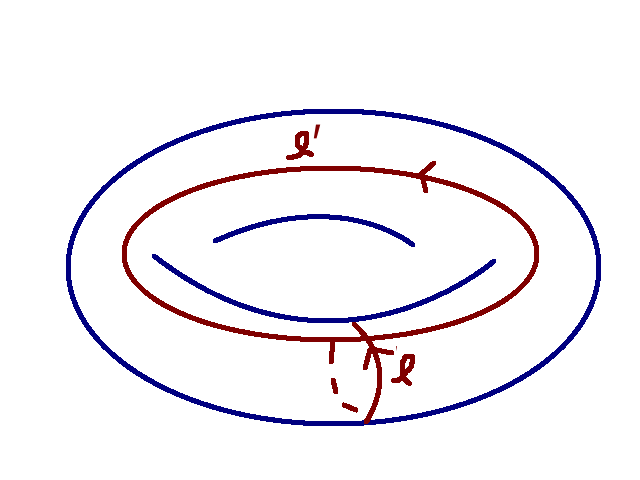}
 \end{center}
\caption{\small Two noncontractible loops on a torus that have a nonzero intersection number $\ell\cap \ell'$.  In the example
shown, the intersection number is 1.}
 \label{Torus3}
\end{figure}

   The basic case -- sketched in the figure -- is that $\ell\cap \ell'=1$.  Moreover,
   we may as well just set $s=s'=1$ since $W_s(\ell)$ is just the $\ell^{th}$ power of $W_1(\ell)$ and similarly for $W_{s'}(\ell')$. 
   If we set $A=W_1(\ell)$, $B=W_1(\ell')$, then the algebra obeyed by $A$ and $B$ is
   \be\label{woof}AB=\exp(2\pi i/r)BA.\ee
     An irreducible representation of this algebra has dimension $r$, 
   because
   \be\label{oof}B\to ABA^{-1}=\exp(2\pi i/r)B\ee
   multiplies any eigenvalue of $B$ by $\exp(2\pi i/r)$.
    So $r$ states are needed to represent this algebra and actually $r$ states are enough.
   These are the $r$ ``ground states of Chern-Simons theory on
   a torus,'' for the case of the gauge group $U(1)$ at ``level'' $r$.

  So this is the basis for the claim that {\it in the limit of a very large system}, a quantum Hall system on a topologically non-trivial manifold
  has a nontrivial vacuum degeneracy.    The condition ``in the limit of a very large system'' is necessary, because the vacuum degeneracy
  is actually slightly lifted by exponentially small effects that result from quasiparticle tunneling around a noncontractible loop.
  Tunneling of a charge $s$ quasiparticle around a noncontractible loop $\ell$ such as that of fig. \ref{Torus2} contributes to the effective Hamiltonian
  a term $W_s(\ell)$ with an exponentially small coefficient.  (This coefficient is definitely not a topological invariant.  It depends on the
  energy of the quasiparticle and the length of the loop $\ell$ -- or more precisely the shortest length of a loop in its homotopy class,
  as this is the most likely tunneling path.)

    \subsection{More On Edge States Of The Integer Quantum Hall Effect}
    
This completes our rather modest introduction to the vast topic of the fractional quantum Hall effect.  
Now we return to the integer quantum Hall effect.
 Largely following  Haldane \cite{Haldane}, we will give a conceptual, noncomputational proof of something of a fact that is familiar from Lecture Two
(another conceptual explanation was already given in section \ref{tcp}): a $2+1$-dimensional system with a Chern-Simons coupling in bulk 
     \be\label{plof}I_\eff=\frac{k}{4\pi}\int \d^3x\,\epsilon^{ijk}A_i\partial_j A_k, \ee
     and $n_+-n_-=k$ chiral edge states on the boundary is completely consistent and anomaly-free.    To do this, we will
     simply describe a physical realization.    First we do this in a continuum language and then we do it via the Haldane model.  
     
  We  couple the field $A$ to a massive $2+1$-dimensional Dirac fermion $\psi$ of charge 1:
  \be\label{utof}I_\psi=\int \d^3x \,\bar\psi\left(\slashed{D}-m\right)\psi.\ee   Since $\psi$ is gapped, we can 
  ``integrate it out'' and get a local
  effective action for $A$ only.    The dominant term at low energies turns out to be
  \be\label{uftof}\frac{\sign m}{2}\ICS(A)=\frac{\sign m}{2}\frac{1}{4\pi}\int\d^3x \epsilon^{ijk}A_i\partial_jA_k. \ee
  The factor of $1/2$ is worrisome as it contradicts gauge invariance.    However, we will always consider combinations in which it is
  absent.    The factor $\sign m$ follows from reflection symmetry (under which $m$ and $\ICS(A)$ are both odd) and dimensional analysis.

    \begin{figure}
 \begin{center}
   \includegraphics[width=2.5in]{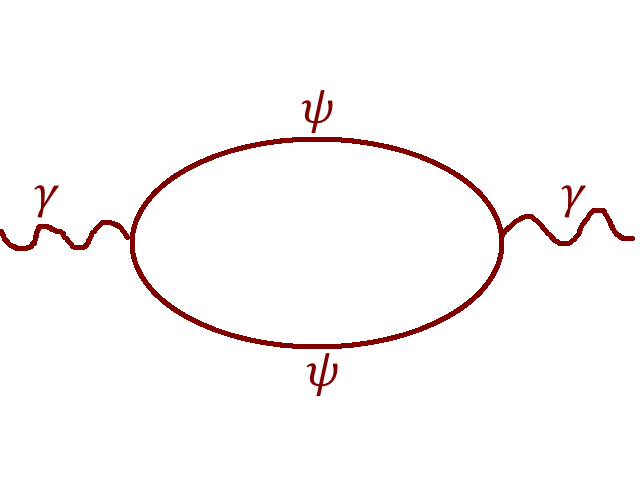}
 \end{center}
\caption{\small The one-loop Feynman diagram associated to the parity anomaly.}
 \label{Feynman}
\end{figure}
  The effective action $(\sign m/2)\ICS(A)$ was first found  \cite{Red,Semenoff}  from a Feynman diagram (fig. \ref{Feynman}), and this is not a difficult calculation.
     However, in the spirit of the present lectures, it is more natural to get this result from the Berry flux.   
  
We know that when any gapped system of charged fermions is ``integrated out,''  the resulting coefficient of $\ICS(A)$  equals
the winding number of the momentum space Hamiltonian. 
The massive Dirac Hamiltonian in $2+1$ dimensions is
\be\label{puft}H=\sigma_x p_x+\sigma_y p_y + m \sigma_z.\ee
  For large $|p|$, the mapping is
\be\label{wuft}(p_x,p_y)\to \left(\frac{p_x}{\sqrt{p_x^2+p_y^2}},\frac{p_y}{\sqrt{p_x^2+p_y^2}},0\right), \ee
which winds around the equator of the sphere (fig. \ref{Equator}).
   \begin{figure}
 \begin{center}
   \includegraphics[width=2.5in]{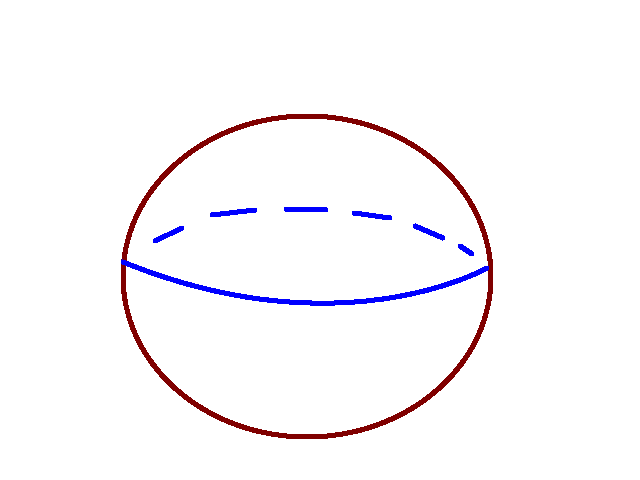}
 \end{center}
\caption{\small The equator of the two-sphere. }
 \label{Equator}
\end{figure}   
    
The full mapping 
 \be\label{PPP}   (p_x,p_y)\to \left(\frac{p_x}{\sqrt{p_x^2+p_y^2+m^2}},\frac{p_y}{\sqrt{p_x^2+p_y^2+m^2}},\frac{m}{\sqrt{p_x^2+p_y^2+m^2}}\right)\ee
 has for its image the upper hemisphere or the lower hemisphere, depending on the sign of $m$ (fig. \ref{Equator2}). 
  So the winding
 number is $\frac{1}{2}\sign \,m$, and that is the Chern-Simons coefficient that we get by integrating out $\psi$.
    \begin{figure}
 \begin{center}
   \includegraphics[width=2.5in]{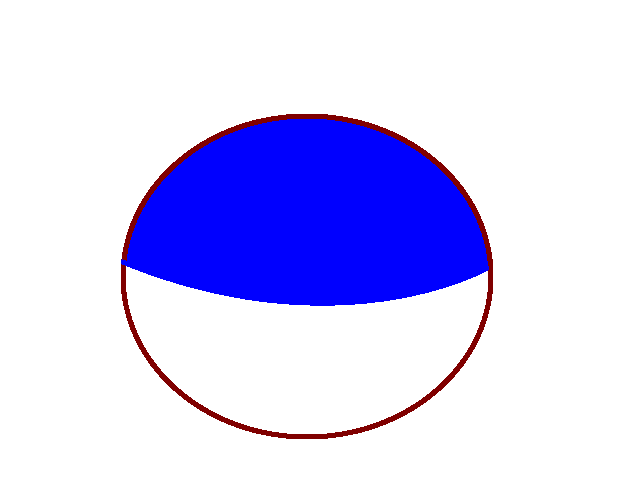}
 \end{center}
\caption{\small The upper hemisphere of the two-sphere.}
 \label{Equator2}
\end{figure}    
 
 For the moment, we want to consider a theory that is gapped and trivial at low energies, so along with $\psi$, we add a second fermion field $\psi'$ of mass
 $-m$.    This combination is trivial in the sense that the total Chern-Simons coefficient obtained by integrating out $\psi$ and $\psi'$ is
 \be\label{quft}\frac{1}{2}\left(\sign\,m+\sign\,(-m)\right)=0.\ee
 There is no induced Chern-Simons coupling, so there is no anomaly on a manifold with boundary.
 
 On a sample with boundary, we want a boundary condition such that the system remains trivial even along the boundary
 -- no edge excitations at all.    This will certainly be consistent and physically sensible!   On a half-space $x_1\geq 0$,
 what boundary condition will ensure that nothing happens along the boundary?  
 
A  boundary condition that does the trick is
 \be\label{plift}\left.\psi'\right|_{x_1=0}=\gamma_1\left.\psi\right|_{x_1=0}.\ee
 Recall that  $\psi$ and $\psi'$ have equal and opposite masses
 \be\label{ulift}\left(\slashed{\partial}-m\right)\psi=0=\left(\slashed{\partial}+m\right)\psi',\ee
 and that in $2+1$ dimensions, the fermion mass is odd under reflection.  
 So if we combine $\psi$ and $\psi'$ to a single fermion $\hat\psi$ defined on all of $\RR^3$ by
 \be\label{nuft}\hat\psi(x_1,x_2;t)=\begin{cases}\psi(x_1,x_2;t) & {\mathrm{if}}~x_1\geq 0 \\
                                       \gamma_1\psi'(-x_1,x_2;t) & {\mathrm{if}}~x_1\leq 0, \end{cases} \ee
   then $\hat\psi$ just obeys
 \be\label{obeys}\left(\slashed{\partial}-m\right)\hat\psi=0,\ee
 and certainly has no gapless mode.
 
 Now, while keeping the fermion kinetic energy and the boundary conditions fixed, we change the sign of the mass of $\psi'$ and
take both  $\psi$ and $\psi'$ to have the same mass $m>0$.  This cannot affect the consistency of the theory since
 the mass is a ``soft'' perturbation.    Of course, when the mass of $\psi'$ passes through zero, the theory becomes ungapped and passes
 through a phase transition.    What is there on the other side of this transition?    The Hall conductivity -- that is
 the coefficient of $\ICS(A)$ in the effective action -- is now
 $\frac{1}{2}\left(1+1\right)=1. $
   By itself this would be anomalous.    But the Dirac equation for $\hat\psi$ is now
      \be\label{gift}\left(\slashed{\partial}-m\,\sign(x_1)\right)\hat\psi=0,\ee and thus the mass of $\hat\psi$ changes sign in passing through
      $x_1=0$.    As in a classic analysis by Jackiw and Rebbi \cite{JR}, this change in sign of the mass leads to the
      existence of a gapless mode supported near $x_1=0$.   The relevant solution is quite similar to what we have already seen in
      eqn. (\ref{ogg}):
      \be\label{zoggg}\psi=\exp(-m|x_1|)\psi_\parallel,\ee
      with 
      \be\label{trogg} \g_1\psi_\parallel =- \psi_\parallel,~~~\slashed{\partial}^\parallel\psi_\parallel=0. \ee
      The condition $\g_1\psi_\parallel=-\psi_\parallel$ determines $\psi_\parallel$ to have definite chirality in the $1+1$-dimensional
      sense, so what we get this way is a chiral edge mode that propagates along the boundary at $x_1=0$.
      
      Thus we have a manifestly consistent construction of a $2+1$-dimensional 
      system that in bulk has an effective action $\ICS(A)$ (plus terms of higher
      dimension) and along the boundary has a chiral edge mode.    Had we started with $k$ pairs $\psi,\psi'$, we would have arrived
      in the same way at a bulk action $k\ICS(A)$ and $k$ chiral edge modes.    So we have confirmed the consistency of this
      combined system without having to investigate the ``anomalies'' of the chiral edge modes.
      
      \subsection{Haldane's Model Of Graphene}\label{hmg}
      
        \begin{figure}
 \begin{center}
   \includegraphics[width=2.5in]{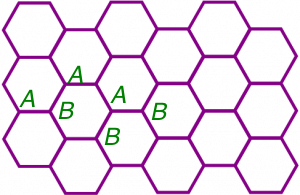}
 \end{center}
\caption{\small  A hexagonal or honeycomb lattice.  This is a bipartite lattice; lattice sites can alternately be labeled $A$ or $B$ in such
a way that the nearest neighbors of $A$ sites are $B$ sites and vice-versa.  The $A$ sites are all equivalent under lattice translations
and the same is true of the $B$ sites.  But there are no translation symmetries exchanging $A$ and $B$
sites.  A unit cell, accordingly, contains one $A$ site and one $B$ site.}
 \label{honey}
\end{figure}
      It remains to describe how Haldane realized this system in a condensed matter model -- a small perturbation of the standard
      band Hamiltonian of graphene.    Graphene is an atomic monolayer of carbon atoms arranged in a hexagonal (or honeycomb) lattice,
      sketched in fig. \ref{honey}. We think
      of this as a lattice in the $x-y$ plane, with $z$ as the normal direction.
      A carbon atom has 6 electrons; 2 of them are in $1s$  states
      and 3 more go into forming covalent bonds with the 3 nearest neighbors of any given atom.  (One can think of the electrons in these
      bonds as hybridized $2s$, $2p_x$, and $2p_y$ electrons.)   We are left with 1 electron per atom, which is going to go into
      the $2p_z$ orbital -- with spin up or down.    Thus the two $2p_z$ orbitals will be ``half-filled.''
      
      The honeycomb lattice has two atoms per unit cell.  Each unit cell has an $A$ atom and a $B$ atom, as explained in the figure.   So the $2p_z$ orbitals form two bands (not counting spin) and we want to
      ``half-fill'' these bands.

              \begin{figure}
 \begin{center}
   \includegraphics[width=2.5in]{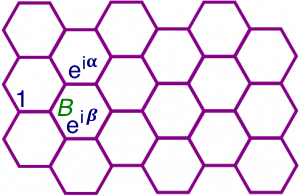}
 \end{center}
\caption{\small   Hopping to a $B$ site from its nearest neighbors.}
 \label{honey2}
\end{figure}
      What happens is dictated by symmetry up to a certain point, but the easiest way to understand it is to first solve a simple model
      \cite{Wallace,Sem,GM}
      in which the Hamiltonian describes ``nearest neighbor hopping'' with amplitude $t$
       from the $A$-lattice to the $B$-lattice and vice-versa.   A shortcut to 
      write the momentum space Hamiltonian is as follows.  
      Pick a point in the $B$ lattice as shown in fig. \ref{honey2}, and let the momentum of an electron (and the normalization of its wavefunction) 
      be such that the amplitudes at the three neighboring $A$ points are $1, e^{i\alpha},
      $ and $e^{i\beta}$.   Here $\alpha$ and $\beta$ are arbitrary angles; they give a convenient parametrization of the Brillouin zone.

      The total hopping amplitude to the indicated $B$ site is then $1+e^{i\alpha}+e^{i\beta}$ (times the hopping constant $t$).    The Hamiltonian is hermitian, so the $B\to A$
      hopping is the complex conjugate of this and the momentum space Hamiltonian is in the $A, B$ basis
      \be\label{hham}H=t\begin{pmatrix} 0  & 1+e^{-i\alpha}+e^{-i\beta}\cr 1+e^{i\alpha}+e^{i\beta}& 0\end{pmatrix}.\ee
.

$H$ is traceless, so a band crossing occurs exactly when there is a zero-mode of $H$.    To find such a zero-mode, we have to solve
\be\label{theq}1+e^{i\alpha}+e^{i\beta} = 0,\ee
with real $\alpha, \beta$.   The equation implies that $e^{i\alpha}$ and $e^{i\beta}$ are complex conjugates, and there are precisely
two solutions
\be\label{omeq}e^{i\alpha}=\frac{1}{2}\left(-1\pm \sqrt{-3}\right)=e^{-i\beta}.\ee
Expanding around either of these solutions, one finds a Dirac-like Hamiltonian, so we have found two ``Dirac points'' in the Brillouin zone.

  \begin{figure}
 \begin{center}
   \includegraphics[width=2.5in]{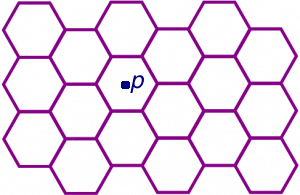}
 \end{center}
\caption{\small  Apart from translation symmetries, the honeycomb lattice has a symmetry group of order 12.  This can be identified as
the group of all symmetries of the lattice that leave fixed a point $p$ at the center of one of the hexagons.  (Up to conjugation by the translation
group, it does not matter which hexagon one picks.)  The lattice has a symmetry group of order 6 consisting of rotations around $p$ by an angle
$2\pi n/6$, $n\in \Z$.  Supplementing this with reflections that leave fixed a line through $p$ gives the group of order 12. }
 \label{Hex5}
\end{figure}

This was a crude model, but the graphene lattice has a lot of symmetries.  Apart from translation symmetries, the symmetries are as follows.   As shown  (fig. \ref{Hex5}), let $p$ be the center of one
 of the hexagons.   Then one can rotate around $p$ by any multiple of $2\pi/6$, and one can also reflect along various axes through $p$.
  For example, one can find a reflection  that maps a given Dirac point to itself -- and therefore (recall the discussion in section \ref{td})
  ensures that the gapless Dirac modes
 that we found in the model remain gapless after any perturbation that preserves the reflection in question.   One can also see that a $2\pi/6$ rotation
 exchanges the two Dirac points, ensuring that they are at the same energy.

Furthermore,
because (in the idealized hopping model) the  band Hamiltonian is traceless
away from the Dirac points, at every momentum away from the Dirac points, precisely one state has energy below the Dirac points
and one has energy above them.  So at half-filling, the Fermi energy is precisely the energy of the Dirac points.  As discussed in section
\ref{bce}, this conclusion remains
valid after any sufficiently small symmetry-preserving perturbations of the ideal Hamiltonian.  It is believed to hold in the real world,
for an ideal graphene crystal  in empty space, with spin-dependent forces turned off (fig. \ref{symb}).

In particular, it is believed that an ideal graphene crystal in the absence of spin-dependent forces has gapless Dirac-like excitations.
 Suitable perturbations involving symmetry breaking and/or spin-dependent forces can give a variety of gapped models.   Haldane
 chose a perturbation that broke some symmetry and gave masses of the same sign to all Dirac modes.   Allowing for spin, this gives a quantum
 Hall coefficient of $2 \times (1/2+1/2)=2.$   Kane and Mele \cite{KM} analyzed the effects of spin-dependent
 forces and arrived at the spin quantum Hall effect, the germ of a 2d topological insulator.

  \begin{figure}
 \begin{center}
   \includegraphics[width=2.5in]{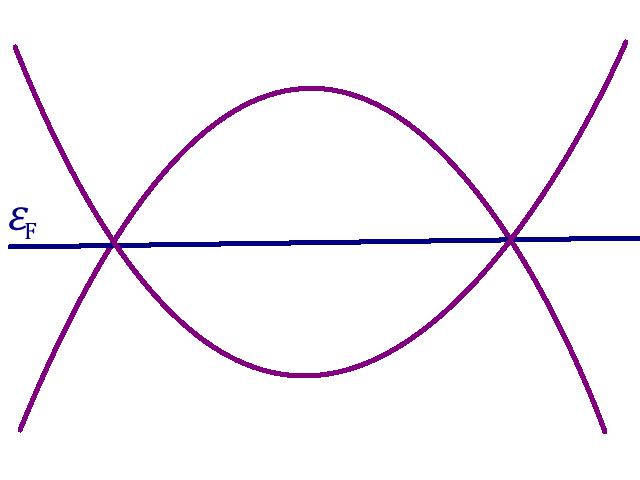}
 \end{center}
\caption{\small  It is believed that the Fermi energy of an ideal graphene crystal in empty space, with spin-orbit
forces turned off, precisely equals the energy of the Dirac points.
}
 \label{symb}
\end{figure}

\end{document}